\documentclass[aps,prd,reprint,preprintnumbers,superscriptaddress,showpacs,twocolumn]{revtex4-1}
\usepackage{latexsym,graphicx,amssymb,amsmath,mathrsfs}
\usepackage{setspace,bm}
\usepackage[breaklinks, colorlinks=true, pdfstartview=FitV, linkcolor=red, citecolor=blue, urlcolor=blue]{hyperref}
\usepackage[usenames]{color}
\usepackage{latexsym}
\usepackage{epstopdf}
\usepackage{mathtools}
\usepackage{url}
\usepackage{comment}
\usepackage{braket}

\allowdisplaybreaks[1]
\usepackage[normalem]{ulem}

\renewcommand\sout{\bgroup \color{red} \ULdepth=-.5ex \ULset}

\begin{document}

\title{
Persistent homology analysis for dense QCD effective model with heavy quarks
}

\author{Kouji Kashiwa}
\email[]{kashiwa@fit.ac.jp}
\affiliation{Fukuoka Institute of Technology, Wajiro, Fukuoka 811-0295,
Japan}

\author{Takehiro Hirakida}
\email[]{hirakida@izumi.ac.jp}
\affiliation{Izumi Chuo high school, Izumi 899-0213, Japan}

\author{Hiroaki Kouno}
\email[]{kounoh@cc.saga-u.ac.jp}
\affiliation{Department of Physics, Saga University, Saga 840-8502,
Japan}

\begin{abstract}
 The isospin chemical potential region is known as the sign-problem free region of quantum chromodynamics (QCD).
 In this paper, we introduce the isospin chemical potential to the three-dimensional three-state Potts model to mimic the dense QCD; e.g., the QCD effective model with heavy quarks at finite density.
 We call it as QCD-like Potts model.
 The QCD-like Potts model does not have the sign problem, but we can expect that it shares some properties with QCD.
 Since we can obtain the non-approximated Potts spin configuration at finite isospin chemical potential where the simple Metropolis algorithm can work, we perform the persistent homology analysis towards exploring the dense spatial structure of QCD.
 We show that the averaged birth-death ratio has the same information with the Polyakov loop, but the maximum birth-death ratio has additional information near the phase transition.
\end{abstract}
\maketitle

\section{Introduction}

Elucidating the phase structure of Quantum Chromodynamics (QCD) at finite temperature ($T$) and real chemical potential ($\mu_\mathrm{R}$) is an interesting and important subject for elementary, nuclear and hadron physics.
Unfortunately, there is the sign problem in QCD at finite $\mu_\mathrm{R}$ and thus we can not have reliable QCD phase diagram yet; for example, see Ref.\,\cite{deForcrand:2010ys} for details of the sign problem.
Several approaches are proposed so far to tackle the sign problem, but the sign problem is not satisfactory resolved in QCD at present.

To sidestep the sign problem, one possibility is employing the QCD effective models.
Since the sign problem is strongly related to the correlation between the gauge field and the chemical potential in the Dirac operator, simplifications of the gauge field dynamics sometimes weaken the sign problem.
For example, the simplest Nambu--Jona-Lasinio (NJL) model~\cite{Nambu:1961tp} does not have the sign problem at finite density.
Of course, such simplification loses some properties of QCD, but we can learn several important lessons from the models.
It should be noted that the sign problem can still remain even if we simplify the gluon dynamics; it depends on how much we keep original QCD nature in the simplified model.

The Potts model with suitably constructed external field is the well known QCD effective model with heavy quarks; for example, see Ref.\,\cite{wu1982potts} for details of the Potts model.
The external field consists of the quark mass and the chemical potential and thus there is the sign problem at finite $\mu_\mathrm{R}$ even if it is weaker than original QCD.
There are several attempts to avoid and/or weaken the sign problem appearing in the Potts model~\cite{deForcrand:2017rfp,Alexandru:2017dcw}.  

There is the analytic expectation that there is the crystalline structure at certain parameter region in the low dimensional Potts model with the external field via the transfer matrix approach; see Ref.\,\cite{Nishimura:2015lit} and references therein.
However, existence of such nontrivial spatial structure is not confirmed yet in the higher dimensional system\,\cite{Akerlund:2016myr}.
The Potts model can be regarded as the QCD effective model with heavy quarks and thus we may expect the non-trivial spatial structure even in QCD if it exists.
In this paper, we try to sidestep the sign problem and investigate the possibility of nontrivial spatial structure from the topological viewpoint.

There are some approaches to weaken the sign problem in QCD and some other theories/models at moderate $\mu_\mathrm{R}$ with low $T$. Famous examples are the complexification of dynamical variables such as the comlpex Langevin dynamics~\cite{Parisi:1980ys,Parisi:1984cs}, the Lefshcetz thimble method~\cite{Cristoforetti:2012su,Fujii:2013sra} and the path optimization method~\cite{Mori:2017pne,Mori:2017nwj} . 
However, those method are not perfect yet and thus it is difficult to apply those methods in whole $\mu_\mathrm{R}$ region.

In this study, we introduce the isospin chemical potential ($\mu_\mathrm{iso}$) to the Potts model and investigate its spatial structure to mimic the dense QCD system with heavy quarks.
The important point is that it is known that the $\mu_\mathrm{R}$ region and the $\mu_\mathrm{iso}$ region are equivalent in the region where the pion condensation does not appear at least in the large $N_\mathrm{c}$ limit; see Ref.\,\cite{Hanada:2011ju} as an example. 
This indicates that we can approximately investigate the $\mu_\mathrm{R}$ region via the $\mu_\mathrm{iso}$ region in the model.
Since we can resolve the sign problem in the Potts model at finite $\mu_\mathrm{iso}$, we here perform the persistent homology analysis~\cite{edelsbrunner2000topological,zomorodian2005computing}.

The persistent homology analysis has been applied to the QCD effective model, i.e., the effective Polyakov-line model~\cite{Hirakida:2018bkf}, and it is expected to be useful to investigate the spatial structure of the theory such as the center cluster structure; for example, see Ref.\,\cite{Gattringer:2010ms,Borsanyi:2010cw,Endrodi:2014yaa} for details of the center cluster.
In this study, we investigate the question that there is the nontrivial spatial structure exist or not at finite density in the QCD-like Potts model.
This study is a first step to investigate the spatial structure of QCD at finite density with the persistent homology. 

This paper is organized as follows.
In the next section \ref{Sec:formalism}, we show the formalism of the Potts model with isospin chemical potential. 
In addition, we briefly explain the persistent homology analysis.
Section \ref{Sec:numerical} shows our numerical results and Sec.\,\ref{Sec:summary} is devoted to the summary.

\section{Formalism}
\label{Sec:formalism}
In this paper, we employ the three-dimensional three-state Potts model with the external field as the QCD effective model with heavy quarks;
see Refs.\,\cite{Alford:2001ug,Kim:2005ck,Kashiwa:2020waa} for details of the relation between the Potts model and QCD.
We here summarize the Potts model and its extension to include the isospin chemical potential.
Some problems for the Potts model with the complexfication of dynamical variables are discussed in appendix\,\ref{sec:ap1}.

\subsection{QCD-like Potts model}

In this subsection, we summarize theoretical insights of the Potts model and its extension.

\subsubsection{Standard Potts model with external field}

The Potts model energy with the external field is given by
\begin{align}
 E &= - \kappa \sum_{{\bf x},{\bf i}}
        \delta_{k_{\mathbf x} k_{{\bf x+i}}}
      - h \sum_{\bf x} k_{\bf x},
\label{eq:energy_ori}
\end{align}
where ${\bf i}$ means the unit vector in the three dimensional space, $\kappa$ denotes the coupling constant, $h$ is the external magnetic field and $k_{\bf x}$ are $\mathbb{Z}_Q$ values which take $0,1,\cdots,Q-1$ at site ${\bf x}$.
This model is so called the $Q$-state Potts model.
In this study, we impose the periodic boundary condition for Potts spins.

If $\kappa$ is positive, the first term in Eq.\,(\ref{eq:energy_ori}) makes spins align to the same direction.
In the case with negative $\kappa$, nearest-neighbor spins favor the different direction.
The last term breaks the $\mathbb{Z}_Q$ symmetry of the system, explicitly.

\subsubsection{Map of chemical potential to external field}
To make the Potts model relate to QCD, we here consider following extension of the Potts energy based on Ref.\,\cite{Alford:2001ug};
\begin{align}
 E &= - \kappa \sum_{{\bf x},{\bf i}}
        \delta_{\Phi_{\mathbf x}\Phi_{{\bf x+i}}}
      - N_\mathrm{f} \sum_{\bf x}
        \Bigl ( h_+ \Phi_{\bf x} + h_- {\bar \Phi}_{\bf x} \Bigr),
\label{Eq:energy}
\end{align}
where $N_\mathrm{f}$ denotes the number of flavor where we assume all favors are degenerated, $h_{\pm}$ is the external field explained in Eq.\,(\ref{Eq:external}) and $\Phi_{\bf x}$ (${\bar \Phi_{\bf x}}$) does the ${\mathbb Z}_3$ values (its conjugate) on each site shown in Eq.\,(\ref{Eq:Polyakov_loop}); it corresponds to the Polyakov-loop in QCD where the number of color is set to $3$.
The color structure is encoded to the functional form of $\Phi$ and ${\bar \Phi}$.
For $N_\mathrm{f}$, we set $2$ in whole computations in this paper.
Since Potts spins are defined on the lattice, we put the space index ${\bf x}$ as the subscript of $\Phi$ unlike the  QCD case (\ref{Eq:QCD}).
For example, the relations between the Potts model and QCD are clearly shown in Refs.\,\cite{deForcrand:2010he,Kim:2005ck,Kashiwa:2020waa}.
This model is sometimes referred as the three-dimensional $\mathbb{Z}_3$ spin model~\cite{Akerlund:2016myr}.

The external fields are consisted of the quark mass and the chemical potential as
\begin{align}
h_\pm &= e^{-\beta(M\mp\mu_\mathrm{R})},
\label{Eq:external}
\end{align}
which is induced from the fermion determinant in QCD partition function; see appendix of Ref.\,\cite{Kashiwa:2020waa} for details.
If the quark mass is sufficiently large in QCD, the three-dimensional three-state Potts model with the external field can be treated as the effective model of QCD at finite $T$ and $\mu_\mathrm{R}$.

In this study, the Potts spin at the site ${\bf x}$ takes so as to $k_{\bf x}=0,1,2$ and then the Polyakov loop and its conjugate are defined as
\begin{align}
    \Phi_{\bf x} &= e^{i 2 \pi k_{\bf x} / 3},~~~~
    {\bar \Phi}_{\bf x} = e^{-i 2 \pi k_{\bf x} / 3}.   
\label{Eq:Polyakov_loop}
\end{align}
Therefore, the spatial averaged values are defined as
\begin{align}
    \Phi = \frac{1}{V} \sum_{\bf x} \Phi_{\bf x},~~~~
    {\bar \Phi} = \frac{1}{V} \sum_{\bf x} {\bar \Phi}_{\bf x},
\end{align}
where $V$ is the spatial lattice volume, $V=L^3$.
It should be noted that this functional form (\ref{Eq:Polyakov_loop}) is just valid in the Potts model:
We can easily understand that the Polyakov loop in the Potts model mimics the Polyakov-loop in QCD by using the setting $\theta_3 = 0$ and $\theta_8 = 2 \pi k/ 3$ with $k=0,1,2$ from Eqs.\,(\ref{Eq:QCD}), (\ref{Eq:QCD2}) and (\ref{Eq:A4}).
Therefore, the Potts model qualitatively picks up the behavior of the Polyakov loop in QCD by using the certain values of $\theta_8$; $\theta_8=2\pi k_{\bf x} /3$.

\subsubsection{Extension to isospin chemical potential}

One possibility to obtain the spin configurations at finite density even if those are approximated configurations, the isospin chemical potential ($\mu_\mathrm{iso}$) is a good candidate; we set $\mu_u = \mu_\mathrm{R}$ and $\mu_d = -\mu_\mathrm{R}$ in the two-flavor ($u$, $d$) system.
In this setting, the quark number density is always zero, but the isospin density can be nonzero.

If the pion and diquark condensates appear, the system with $\mu_\mathrm{iso}$ shows the difference with that with $\mu_\mathrm{R}$, but the systems share similar properties when they do not appear in the systems.
In this paper, we are interested in the system with the heavy quarks and thus both condensations can be neglected.
Therefore, we here employ the system with $\mu_\mathrm{iso}$ to investigate the spatial structure of the spin configuration of Potts model at finite density.
The Potts energy with $\mu_\mathrm{iso}$ is then given by
\begin{align}
 E_\mathrm{iso}
   &= - \kappa \sum_{{\bf x},{\bf i}}
      \delta_{\Phi_{\mathbf x}\Phi_{{\bf x+i}}}
   \nonumber\\
   &  - \sum_{\bf x} \Bigl[
         (h_+ \Phi_{\bf x} + h_- {\bar \Phi}_{\bf x}) + (h_- \Phi_{\bf x} + h_+ {\bar \Phi}_{\bf x}) \Bigr]
  \nonumber\\
   &= - \kappa \sum_{{\bf x},{\bf i}}
      \delta_{\Phi_{\mathbf x}\Phi_{{\bf x+i}}}
  \nonumber\\
   &  - \sum_{\bf x} \Bigl[
         h_+ (\Phi_{\bf x}+{\bar \Phi}_{\bf x}) + h_- (\Phi_{\bf x} + {\bar \Phi}_{\bf x}) \Bigr]
  \nonumber\\
   &= - \kappa \sum_{{\bf x},{\bf i}}
      \delta_{\Phi_{\mathbf x}\Phi_{{\bf x+i}}}
      - N_\mathrm{f}\sum_{\bf x} \Bigl[
         (h_+ + h_-) \cos (\Phi_{\bf x}) \Bigr]
   \nonumber\\
   &\in \mathbb{R}.
\end{align}
Since the external field term facilitates $k=0$ for the spin configuration, it mimics the behavior of ${\Phi}={\bar \Phi} \to 1$ with $\mu_\mathrm{R} \to \infty$.
Of course, $\Phi = {\bar \Phi} \to 0$ with $\kappa \to 0$ and $\mu_\mathrm{R} \to 0$ is naturally obtained.
Therefore, we have these desirable behaviors from the isospin chemical potential in the Potts model, and these are sharp contrast with the complexfied Potts model discussed in appendix.  

In the Potts model, we do not introduce the chiral, diquark and also the pion condensation and thus the $\mu_\mathrm{R}$ and $\mu_\mathrm{iso}$ regions are expected to be almost consistent with each other based on the knowledge obtained in the orbiforld equivalence; for example, see Ref.~\cite{Hanada:2011ju}.     
This fact indicates that we can approximately obtain the Potts spin configuration with $\mu_\mathrm{R}$ via the $\mu_\mathrm{iso}$ region and then we can investigate the dense spatial structure, approximately.

\subsection{Observables}

Most simple quantity which can clarify the system change is the spatial averaged Polyakov-loop defined as
\begin{align}
    \langle \Phi \rangle &= \Bigl\langle \frac{1}{V} \sum_{\bf x} \Phi_{\bf x} \Bigl\rangle,
\end{align}
where $\langle \cdots \rangle$ means the configuration averaged value.
This quantity can represent bulk properties of the system, but it can not see the spatial nontrivial structure.

To investigate the spatial structure of the system, one possible quantity is the spatial correlator of the Polyakov loop which may be fitted as
\begin{align}
    \langle \Phi_0 \Phi^\dag _{r} \rangle
    \sim e^{-m_\mathrm{R} r} \cos(m_\mathrm{I} r),
\end{align}
where $\Phi_0$ means the Polyakov loop at the origin, $m_\mathrm{R,I}$ can be interpreted as the real and the imaginary part of the effective mass of Potts spins and $r$ is the Euclidean distance between two Polyakov-loop operators, $r=\sqrt{x^2+y^2+z^2}$.
At the critical point, the form of the fitting function is slightly modified.
However, the rotational symmetry is explicitly broken by the lattice discretization and also we do not know the actual functional form of the oscillating mode, we consider the correlator for the $x$-direction in this paper.
From the spatial correlator, we  can investigate the simple spatial structures of the system such as the periodic spatial oscillation (inhomogeneity); see Ref.\,\cite{Lenz:2020bxk} for the case of the $1+1$ dimensional Gross-Neveu model as an example.
Actually, we calculate $\langle f(\Phi_0) f(\Phi _{x}) \rangle$ where $f(A)$ mean $\mathrm{Re}$ or $\mathrm{Im}$ of $A$.
Unfortunately, these quantities are usually noisy and thus we need good statistics.

In addition to above quantities, we consider the persistent homology to investigate whether the complicated spatial structure of the system or not.
Some details of the persistent homology are explained in the next section\,\ref{sec:ph}.
Good point is that this quantity is less noisy than the spatial correlator since it is directly related with the topological structure of data.

\subsection{Persistent homology analysis}
\label{sec:ph}

In this study, we employ the persistent homology to investigate the spatial structure of the model at finite density: It is known that the persistent homology can clarify spatial nontrivial structures~\cite{nakamura2015persistent,hiraoka2016hierarchical,donato2016persistent} and also the hidden order~\cite{olsthoorn2020finding}.
In the numerical estimation of the persistent homology, we employ the homcloud~\cite{HomCloud}.
Since the persistent homology is a complicated mathematical concept, we here only show the brief explanation.
If the reader is interested in mathematical details of the persistent homology, see the text and conceptual figures in Sec. II of Ref.\,\cite{Hirakida:2018bkf} and references therein.

\subsubsection{Setting of data space}
In the QCD-like Potts model, there are three independent directions of the spin degree of freedom for each site.
Let us consider the data set $A, B$ and $C$; we save the spatial structure for the data sets.
For sites with $k=0$ spin, corresponding data (coordinate data) are "ON" (occupied) in data set $A$, but other sites in the data set $A$ are "OFF" (empty) if $k \neq 0$ spins are realized.
Via the setting, we can save the spatial structure for $k=0$ spins into the data set $A$.
Similarly, we can save the spatial structure for $k=1$ and $k=2$ spins into the data set $B$ and $C$, respectively. 

In the case of the Polyakov-line model analyzed in Ref.\,\cite{Hirakida:2018bkf}, we need to divide the Polyakov-loop into three domain based on the center symmetry and construct the data set $A,B$ and $C$.
In the QCD-like Potts model, however, we can simply divide the data space into three data sets because the Potts spins are discrete ${\mathbb Z}_3$ quantities by definition.

\subsubsection{Birth and death times of holes}
After setting the data set $A, B$ and $C$, we can consider filtration for each data point and the filtration leads to the persistent homology.
In the persistent homology analysis, we consider the $r$-ball model.
The ball has the radius $r$ for each data point which is corresponding to occupied sites. 
The center of each ball is set to the position of each site.
The value of $r$ is controlled by the fictitious time $t$.
When $r$ is sufficiently small ($t$ is small), each balls are isolated, but
neighborhood balls are overlapped when $r$ becomes large ($t$ is large).
Therefore, there should be the time where the hole appears for the overlapped balls; this time is the so called {\it birth time}, $t_B$.
Of course, such hole will be vanished when $r$ becomes sufficiently large; this time is the so called {\it death time}, $t_D$.
Unfortunately, construction of $r$-balls is numerically difficult and thus we employ the alpha complex to approximate the $r$-ball model; see Ref.\,\cite{Hirakida:2018bkf} for details of the alpha complex as an example.
Then, the birth and death times are related to the squared radius of the ball, $r^2$.

\begin{figure}[t]
 \centering
 \includegraphics[width=0.235\textwidth]{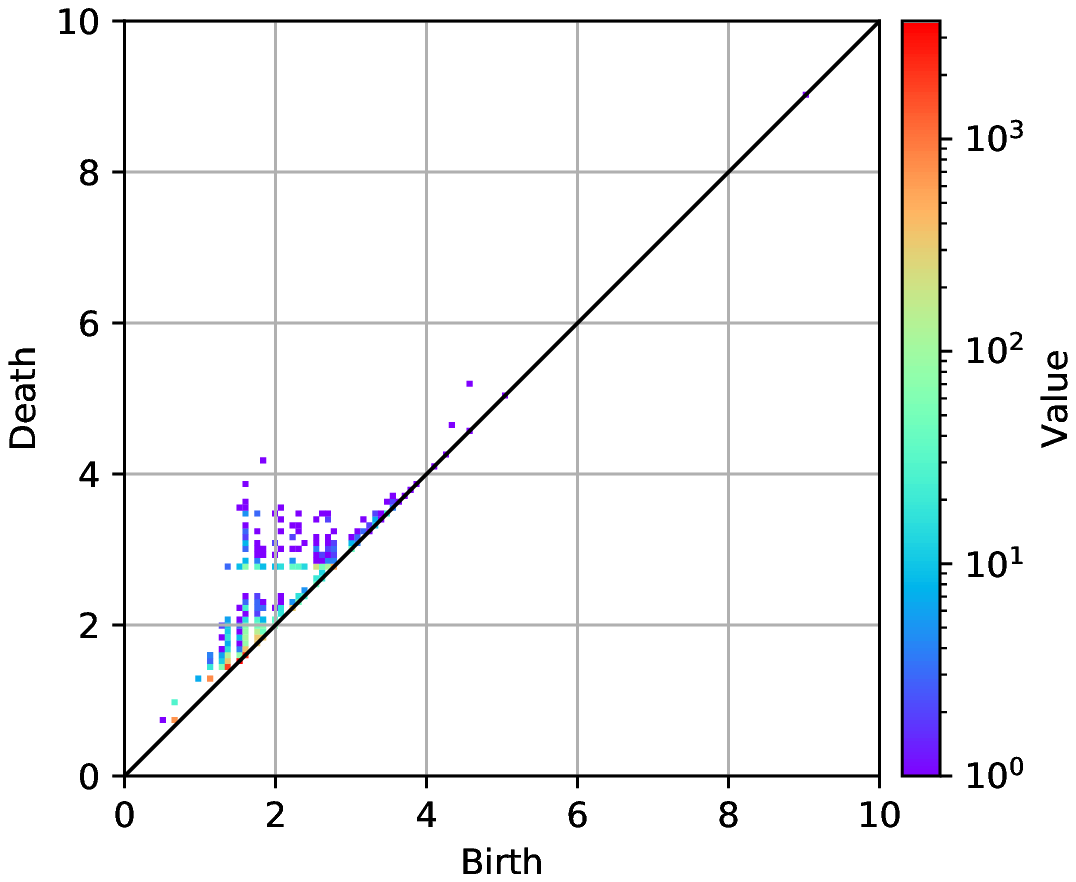}
 \includegraphics[width=0.235\textwidth]{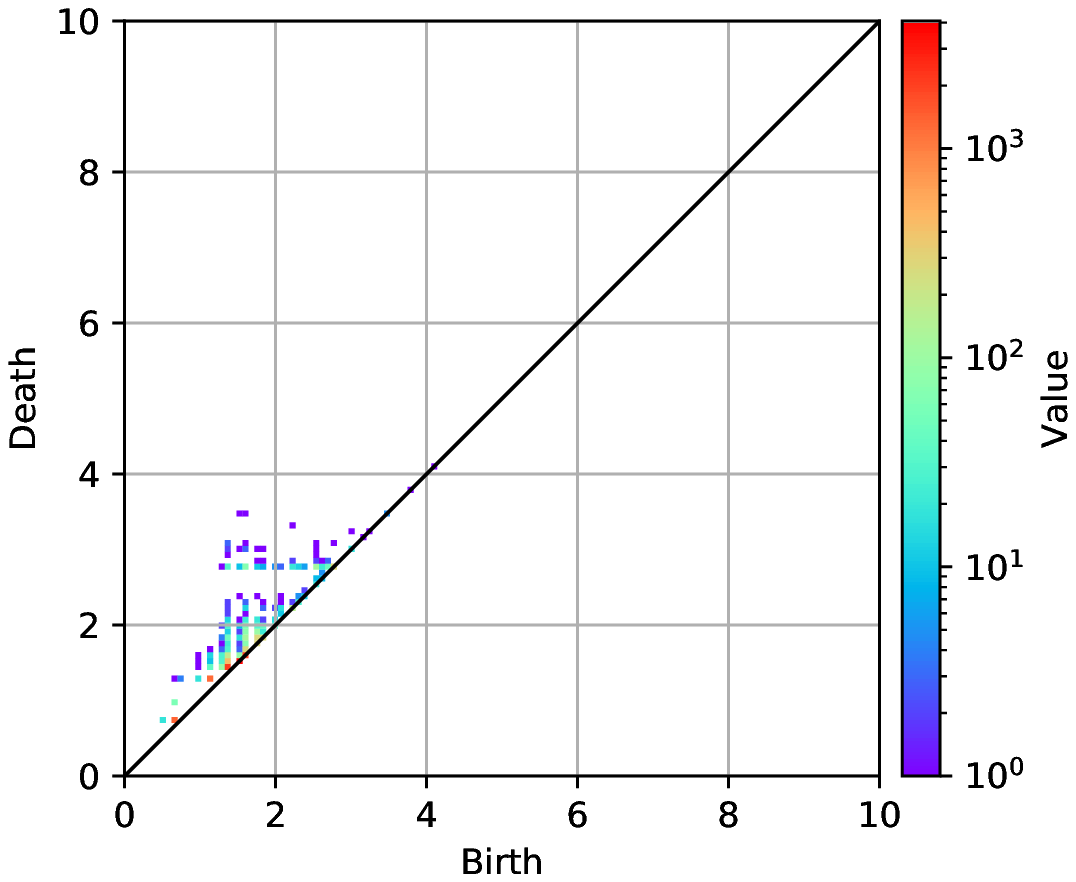}
 \includegraphics[width=0.235\textwidth]{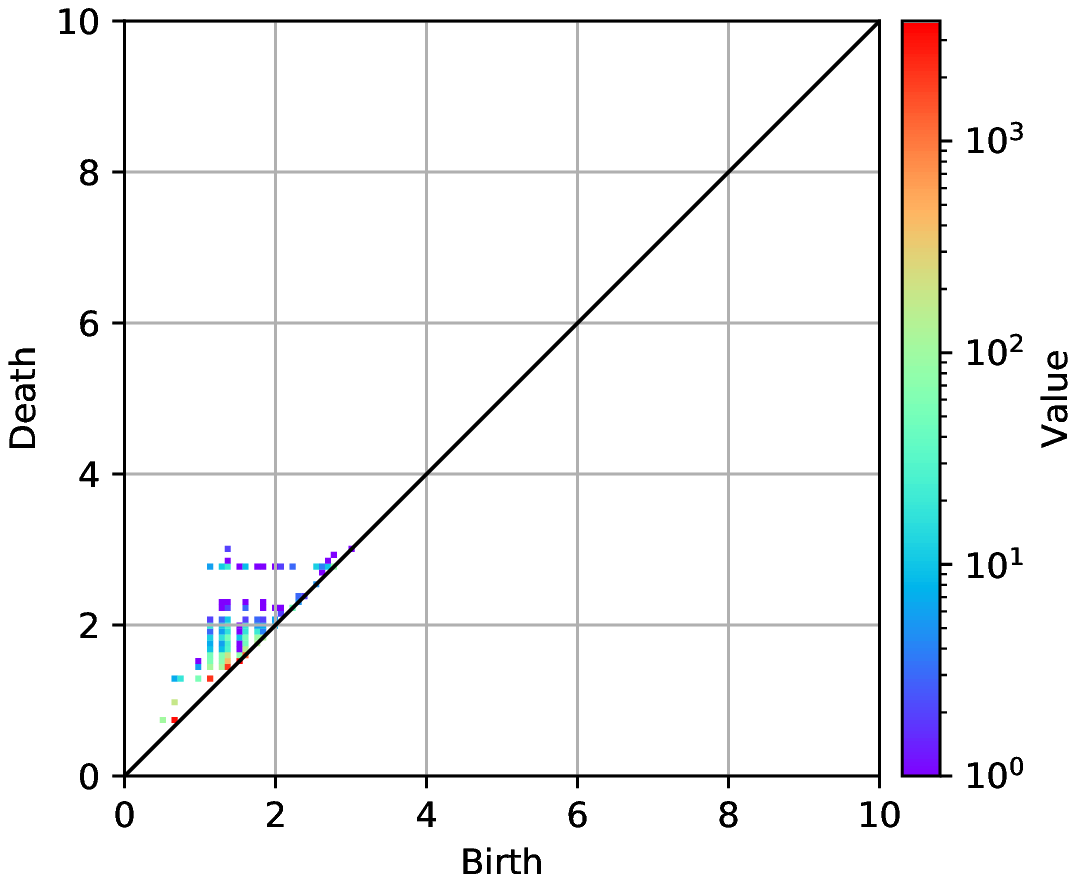}
 \includegraphics[width=0.235\textwidth]{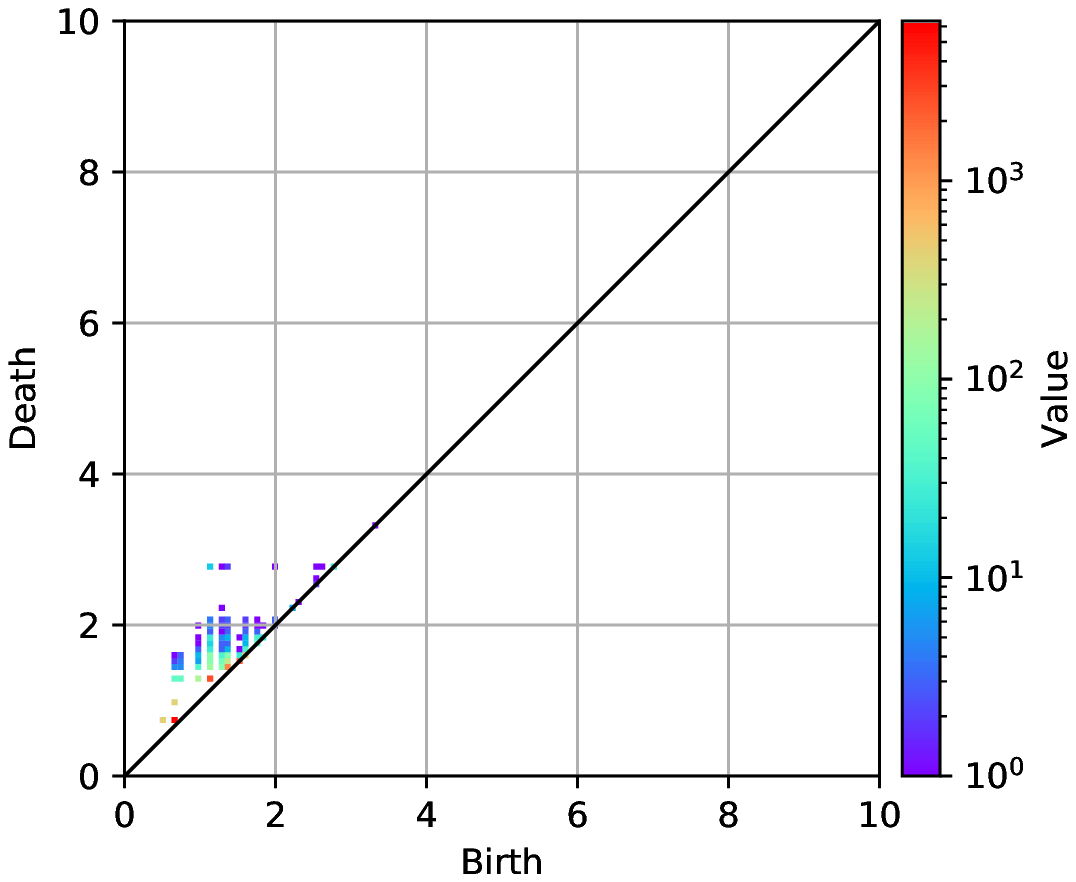}
 \caption{The persistent diagram for randomly distributed data in the $30^3$ squared lattice system as a function of the birth and death times.
 The occupation ratio of the system is about $33\%, 40\%, 50\%$ and $60\%$ from the top-left $\to$ top-right $\to$ left bottom $\to$ right-bottom panels, respectively.
 The legend, Value, means the number of data which appear at the same point.}
\label{fig:ph-r-u}
\end{figure}
Figure \ref{fig:ph-r-u} show the persistent diagram for randomly distributed data as an example.
Here, about $33\%, 40\%, 50\%$ and $60\%$ of sites are occupied from the top-left to the right-bottom panel.
The situation with the $33\%$ occupation is corresponding to the ideal confined phase where nontrivial spatial structures are absent and thus deviations from the randomly distributed case should be related with the confinement-deconfinement nature and also the spatial nontrivial structure.
In the case of the uniform distributed data which means that all sites are occupied, there are only the trivial holes; the actual value of the birth and death times for the trivial holes are $t_\mathrm{B} = 0.5$ and $t_\mathrm{D} = 0.75$, respectively.
Those trivial holes can be easily imaged from the system where all sites are occupied; the system is consist of smallest cubes.
This situation is corresponding to the ideal deconfined phase for $k=0$ spins in this lattice model.
Dominant hole structures are clarified in Ref.\,\cite{Hirakida:2018bkf} in the confined phase at low $\kappa$ and zero density. 

\subsubsection{Ratio of birth and death times}
Since the persistent diagram which is the two-dimensional diagram is not convenient in the lattice simulation because we must take the configuration average and then the two-dimensional diagram is obtained for each configuration.
To simply visualize the persistent diagram, the averaged ratio of the birth and death times has been proposed in Ref.\,\cite{Hirakida:2018bkf};
\begin{align}
    R &= \Bigl\langle \frac{1}{N_h} \sum_{i} \frac{t_{D,i}}{t_{B,i}} \Bigr\rangle,
\label{eq:ratio}
\end{align}
where $N_\mathrm{h}$ is number of all possible holes and $\sum_i$ means that we sum up $t_{D}/t_{B}$ for all possible holes.
In this study, we show this ratio in addition to the standard persistent diagram.

However, the simple definition of the average ratio (\ref{eq:ratio}) may not be good because relatively trivial holes which appear near the diagonal line on the persistent diagram can dominate $R$; it is known that distant data from the diagonal line usually have the meaningful nontrivial spatial structures.
Therefore, we will investigate the maximum ratio for all possible holes in Sec.\,\ref{Sec:numerical} in addition to the above averaged ratio.
Such maximum ratio is expected to be relevant for the nontrivial large spatial structure.

\section{Numerical results}
\label{Sec:numerical}

\begin{figure}[t]
 \centering
 \includegraphics[width=0.23\textwidth]{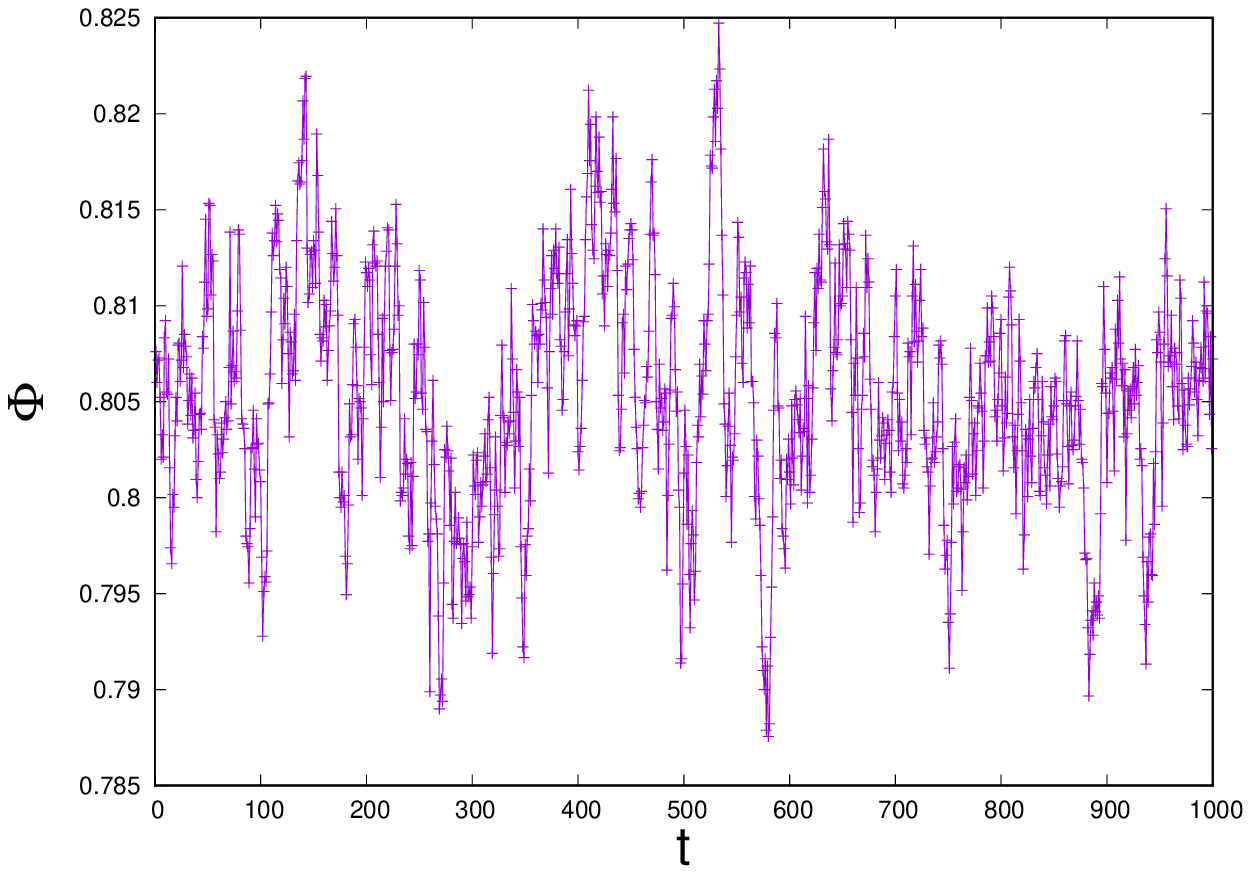}
 \includegraphics[width=0.23\textwidth]{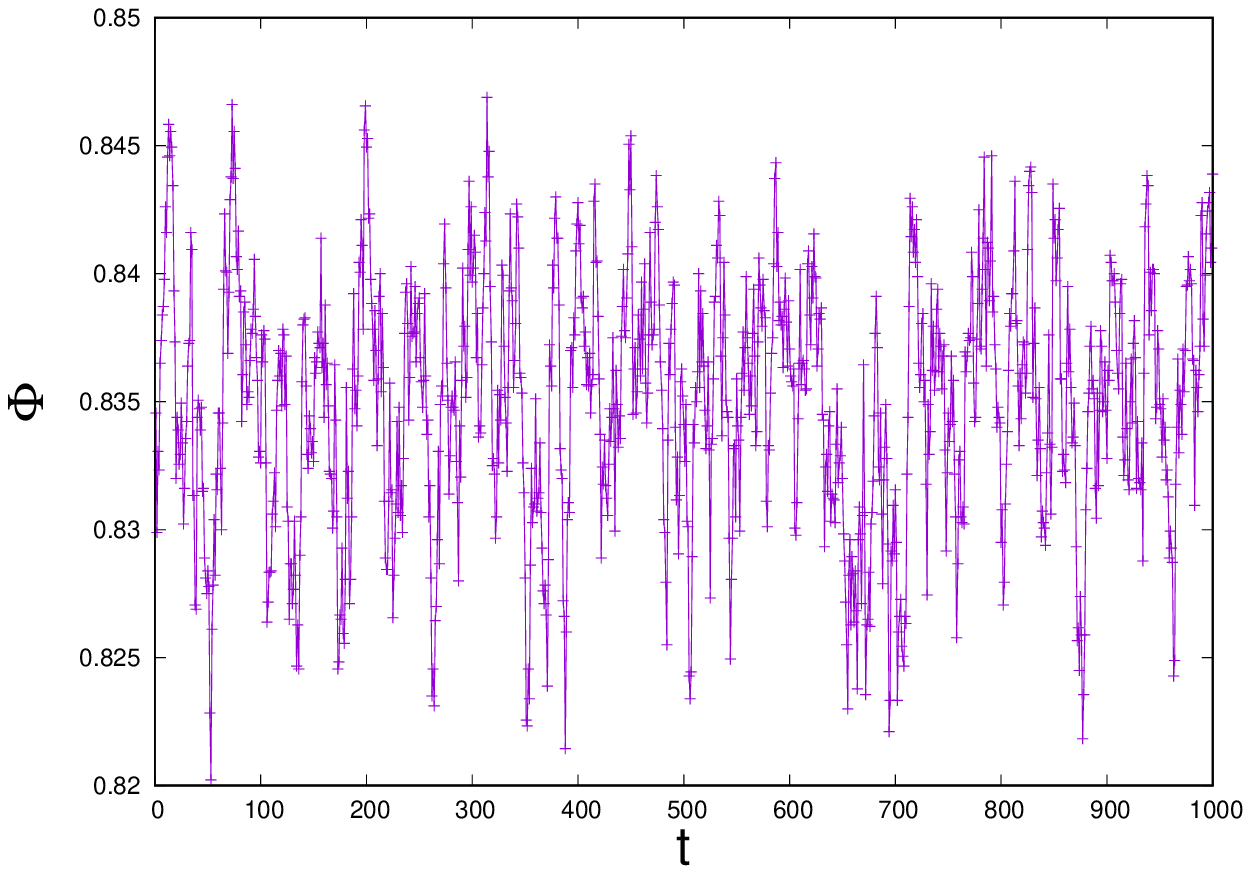}
 \caption{
 The Monte-Carlo evolution of the spatial averaged Polyakov-loop after thermalization.
 The left (right) panel is the result with $\kappa=0.6$ and $\mu_\mathrm{iso}=2$ ($\kappa=0.6$ and $\mu_\mathrm{iso}=6$).
 Each symbol is the result with corresponding configuration which is obtained via the standard Metropolis algorithm.
 The horizontal axis $t$ means the label number of configurations.
 }
\label{fig:evolution}
\end{figure}
In this study, we use the $V=L^3=30^3$ squared lattice system.
we generates $10^3$ configurations for each $L^3$ updation ($1$ Monte-Carlo step) after the thermalization by using the standard Metropolis algorithm; the transition probability (${\cal P}$) for the acceptance or rejection of the single spin flip is defined as ${\cal P} = \mathrm{min} [1,\exp(-\beta \Delta E_\mathrm{iso})]$ where $\beta$ is the inverse temperature $\beta = 1/T$ and $\Delta E_\mathrm{iso}$ is the energy difference about the single spin flip.  
The lattice spacing is set to $a=1$.
Statistical errors are estimated by using the Jackknife method.
For the random number generation, we employ the Mersenne Twister algorithm~\cite{matsumoto1998mersenne}.
The mass parameter $M$ and temperature $T$ are set to $10$ and $1$ in whole calculation, respectively.

Figure \ref{fig:evolution} shows the Monte-Carlo evolution of the spatial averaged Polyakov-loop.
The left (right) panel is the result with $\kappa = 0.6$ and $\mu_\mathrm{iso}=2$
 ($\kappa = 0.6$ and $\mu = 6$) after thermalization as a typical example.
This figure means that we can well generate configurations even at finite $\mu_\mathrm{iso}$.

\subsection{Basic phase structure}

\begin{figure}[t]
 \centering
 \includegraphics[width=0.34\textwidth]{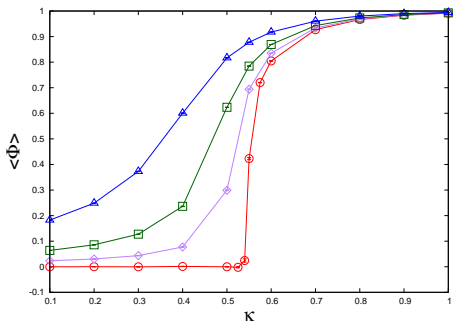}
 \caption{
 The $\kappa$-dependence of the Polyakov loop.
 The open circle, diamond, square and triangle symbols are results with $\mu_\mathrm{iso}=0, 6, 7$ and $8$, respectively.
 Lines are just eye guides.
 }
\label{fig:p-t}
\end{figure}

Figure~\ref{fig:p-t} shows the $\kappa$-dependence of the Polyakov loop with $\mu_\mathrm{iso}=0, 6, 7$ and $8$.
Statistical errors are small and thus error-bars are in the symbols.
Because of the finite size effect, the phase transition is smeared if they exist in the thermodynamic limit. 
In the Potts model, $\kappa$ is treated as the temperature and thus we can see the increasing behavior of $\langle \Phi \rangle$ with increasing $\kappa$.
In addition, $\langle \Phi \rangle$ increases with increasing $\mu_\mathrm{iso}$ as we expected:
Since $\mu_\mathrm{iso}$ enhances the explicit $\mathbb{Z}_3$ symmetry breaking and it leads nonzero $\langle \Phi \rangle$, the first-order thermal phase transition is weaken when $\mu_\mathrm{iso}$ becomes large.
In the present model, statistical errors are under controlled for each $\mu_\mathrm{iso}$.
Therefore, we can think that the present QCD-like Potts model shares several properties with thermal and dense QCD matter with heavy quarks and thus it is a convenient model in this study.
It is noted again that the QCD-like Potts model does not have the sign problem and thus we can have exact Potts spin configurations.

\begin{figure}[t]
 \centering
 \includegraphics[width=0.38\textwidth]{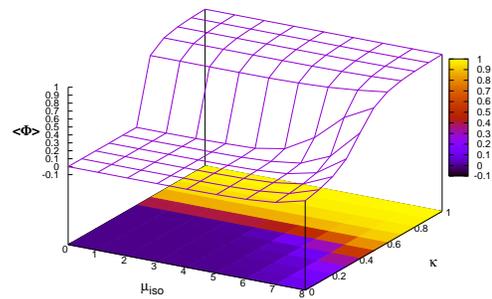}
 \caption{
 The mean value of Polyakov loop on the $\mu_\mathrm{iso}$-$\kappa$ plane.
 Statistical errors are small and thus we do not show them here.
 }
\label{fig:con}
\end{figure}

Figure \ref{fig:con} shows the mean value of Polyakov loop on the $\mu_\mathrm{iso}$-$\kappa$ plane.
Statistical errors estimated by using the Jackknife method are quite small and thus we do not show it here.
We can clearly see the $\kappa$- and the $\mu_\mathrm{iso}$-dependence of $\langle \Phi \rangle$ from the figure, and the behavior is well matched with our expectation in dense QCD with heavy quarks.
At low $\mu_\mathrm{iso}$, there is the first-order thermal transition, but not at large $\mu_\mathrm{iso}$ because the $\mu_\mathrm{iso}$ contribution breaks the $\mathbb{Z}_3$ symmetry explicitly as explained above.
This indicates that there should be the second-order transition point which is the critical endpoint at finite $\kappa$ and $\mu_\mathrm{iso}$.
From the next subsection, we discuss the spatial structure at finite $\mu_\mathrm{iso}$.

\subsection{Spatial structure}

\begin{figure}[t]
 \centering
 \includegraphics[width=0.18\textwidth]{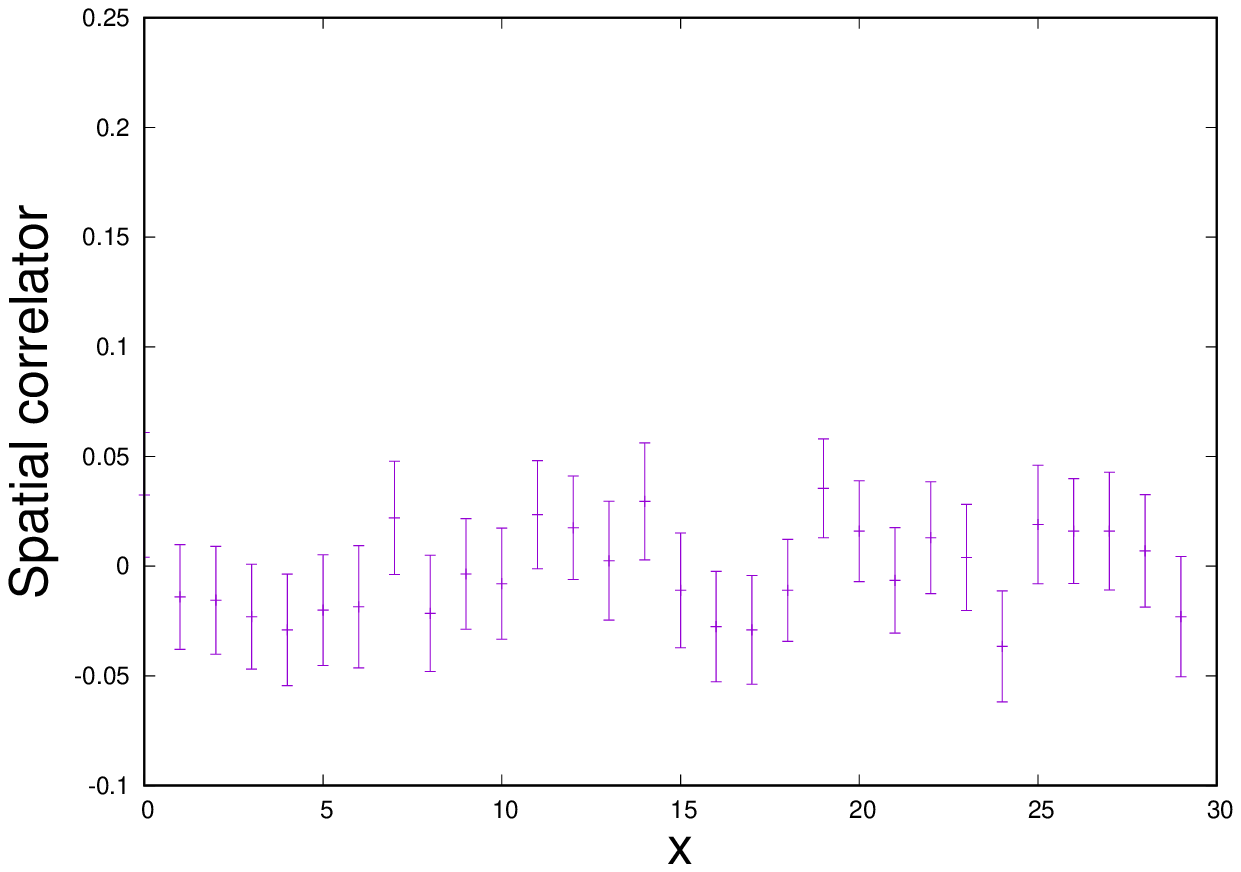}
 \hspace{4mm}
 \includegraphics[width=0.18\textwidth]{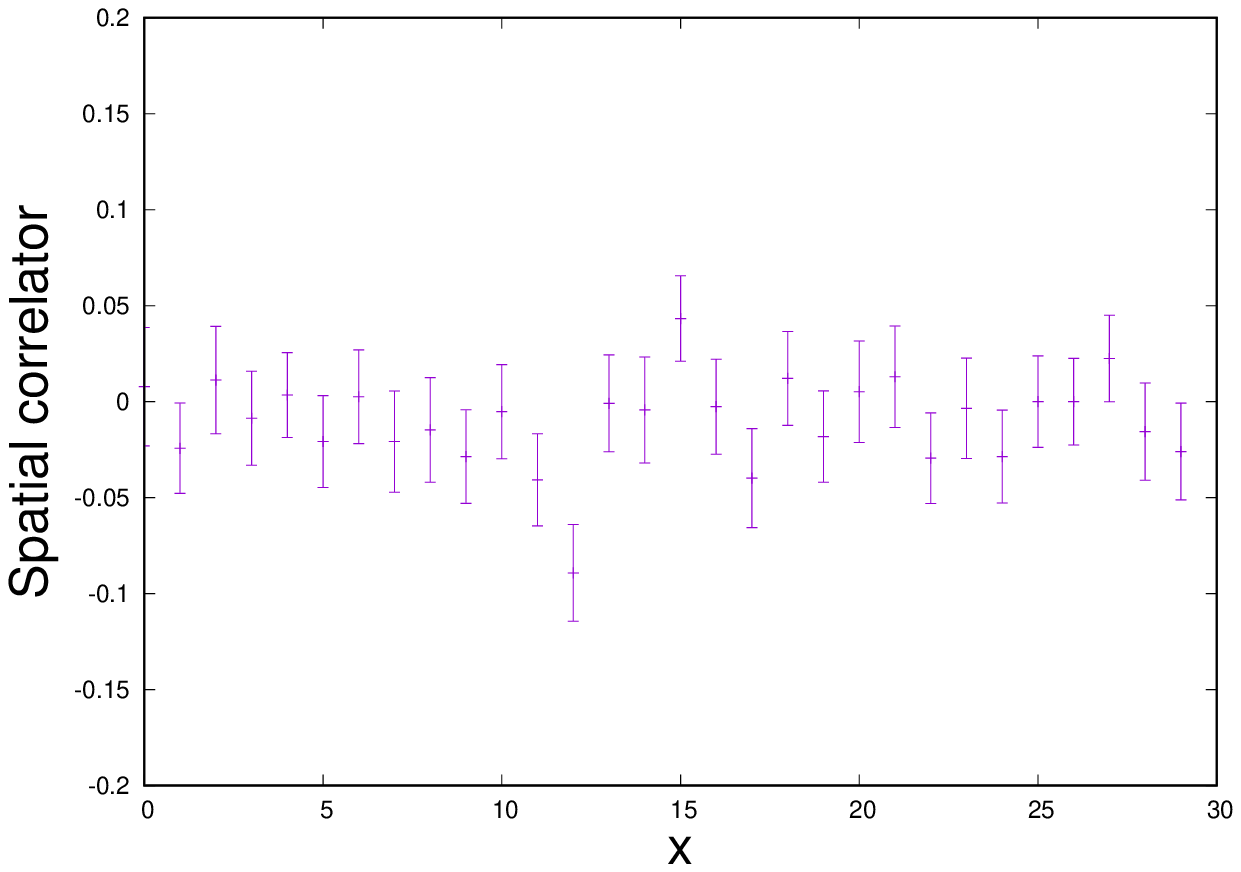}\\
 \includegraphics[width=0.18\textwidth]{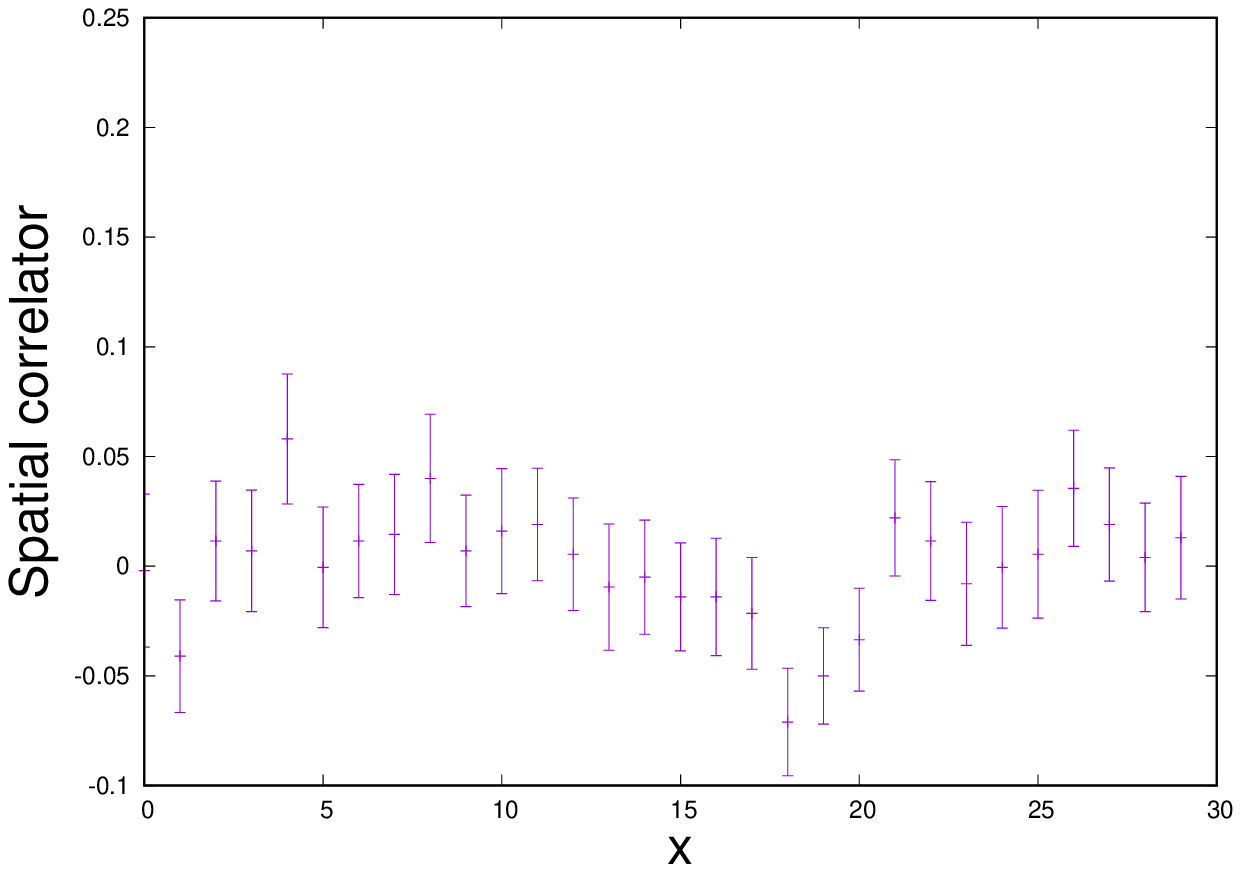}
 \hspace{4mm}
 \includegraphics[width=0.18\textwidth]{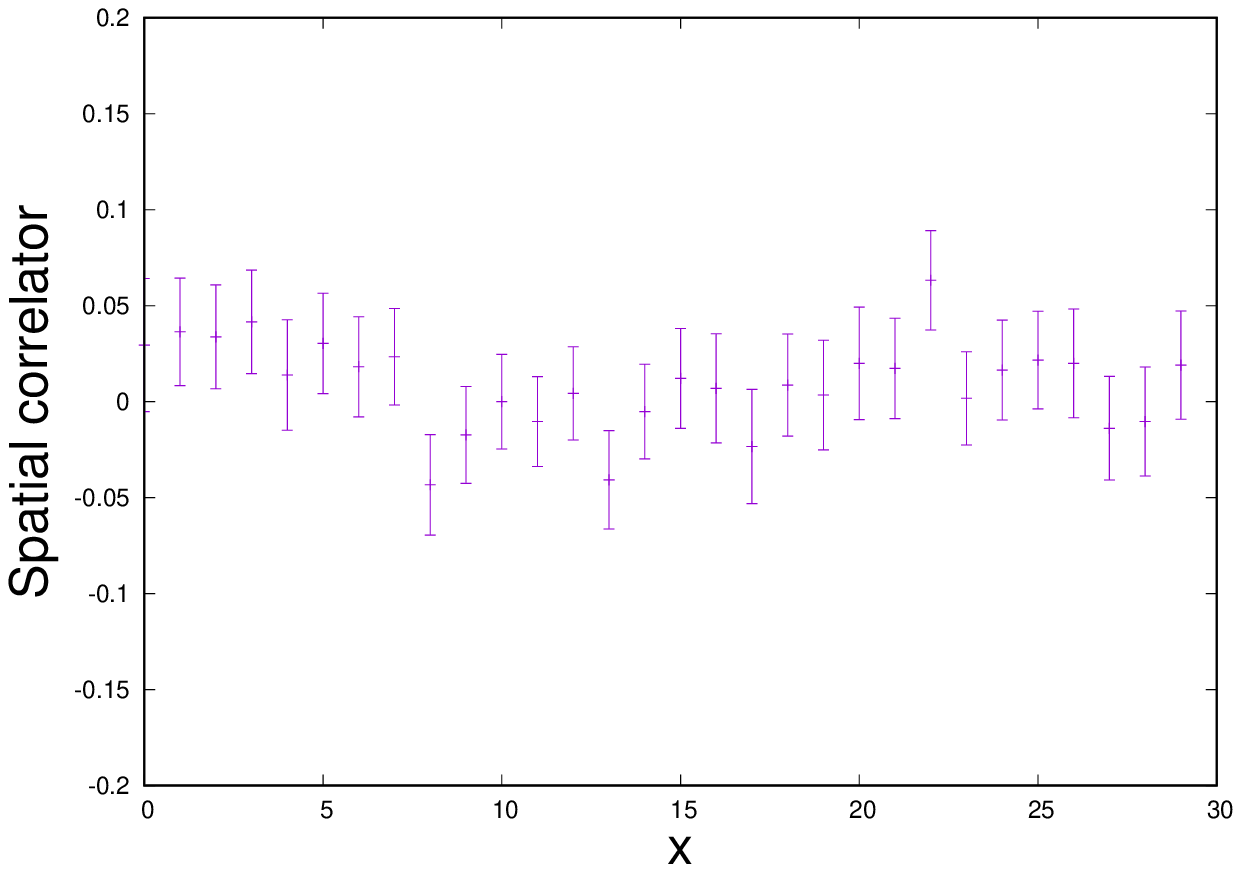}\\
 \includegraphics[width=0.18\textwidth]{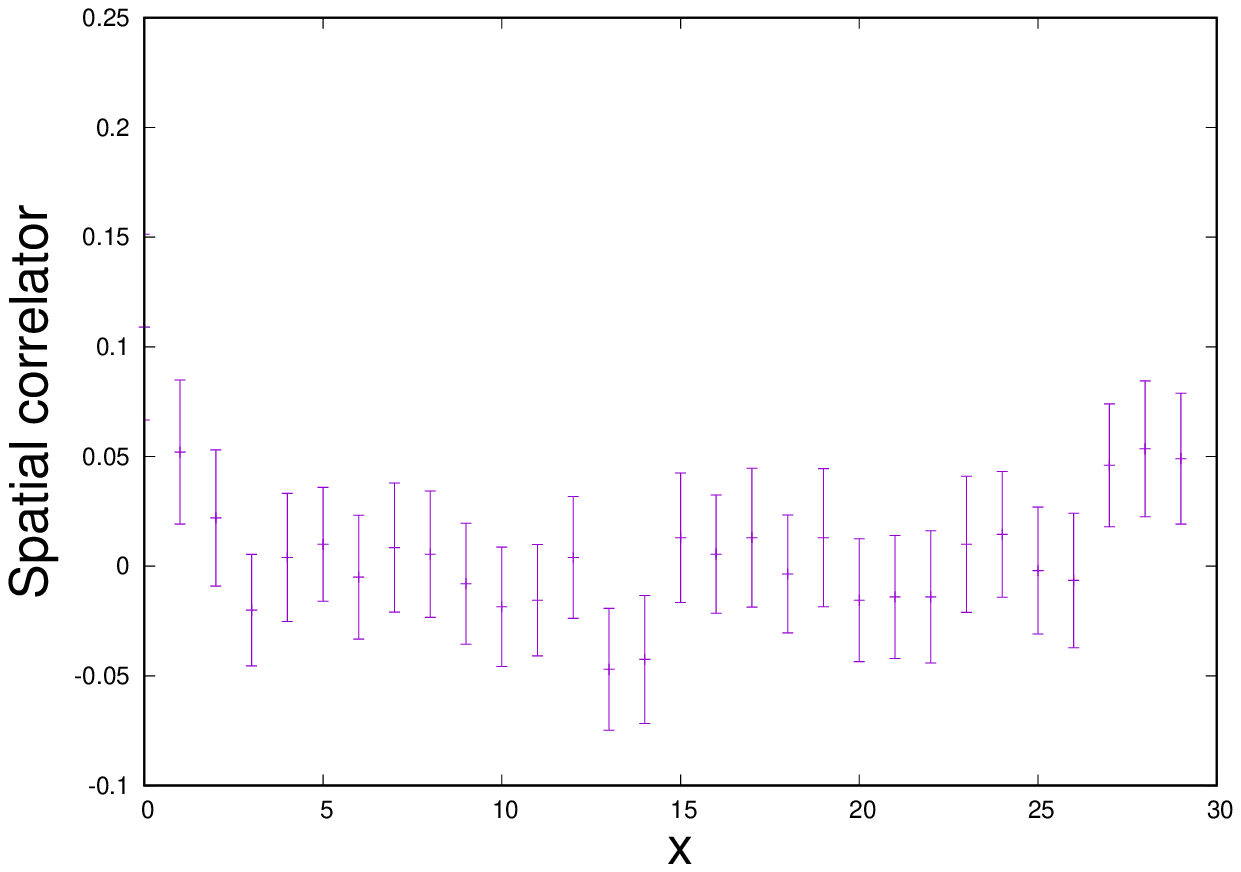}
 \hspace{4mm}
 \includegraphics[width=0.18\textwidth]{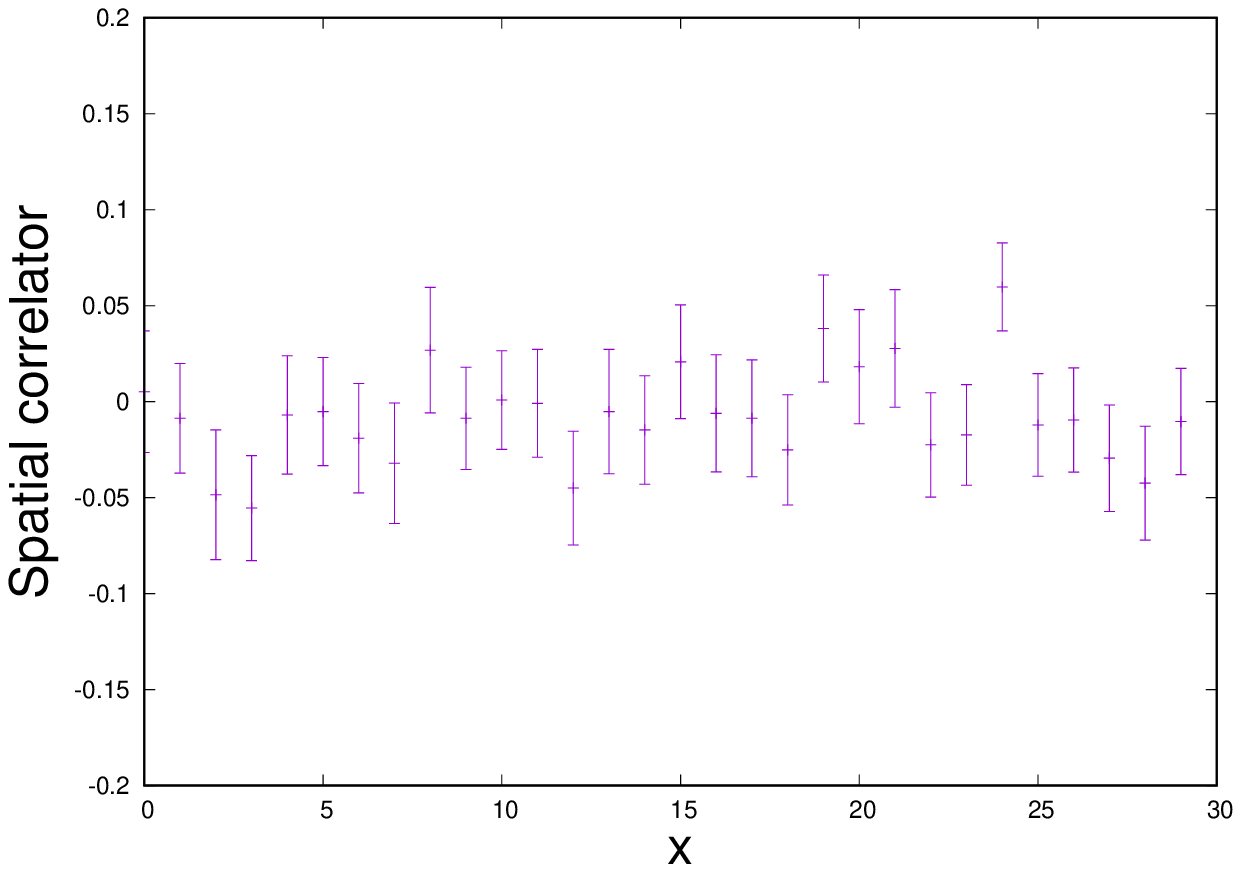}\\
 \includegraphics[width=0.18\textwidth]{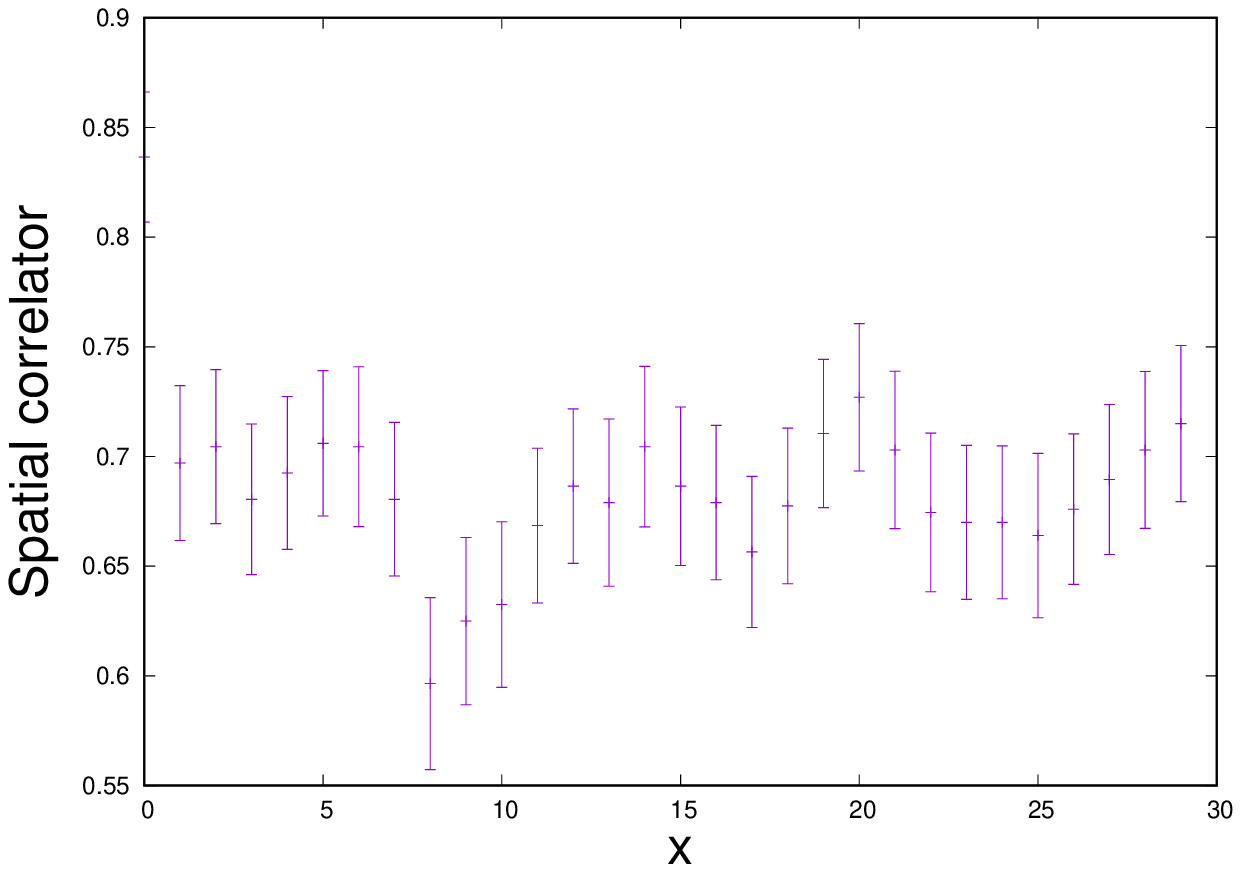}
 \hspace{4mm}
 \includegraphics[width=0.18\textwidth]{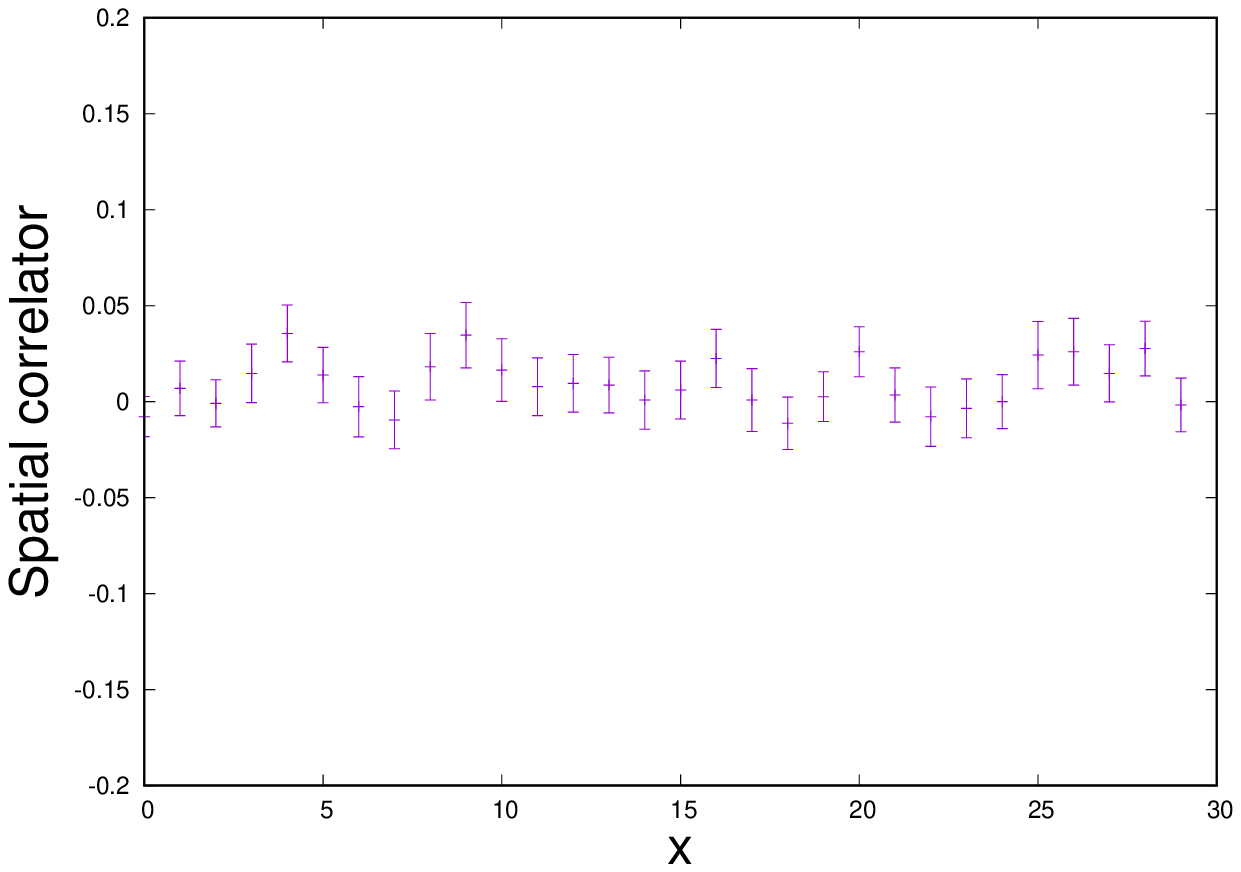}\\
 \caption{
 The spatial correlators for the $x$-direction at $\mu_\mathrm{iso}=5$ with $\kappa=0.3, 0.4, 0.5$ and $0.6$ from the top to the bottom panel, respectively.
 The left and right panels show the spatial correlators for the real and imaginary parts of the Polyakov loop, respectively.
 }
\label{fig:correlator}
\end{figure}
In this section, we investigate the spatial structure at finite $\mu_\mathrm{iso}$ by using the spatial correlators and the persistent homology.
Figure \ref{fig:correlator} shows the spatial correlators for $\mathrm{Re}\,\Phi$ and $\mathrm{Im}\,\Phi$ at $\mu_\mathrm{iso}=5$ with $\kappa=0.3, 0.4, 0.5$ and $0.6$ from the top to the bottom panel.
Left (right) panels are result of the spatial correlator for the real (imaginary) part of the Polyakov loop.
The spatial correlators are very noisy comparing with the Polyakov loop with the present statistics. 
From the figures, we can not see clear tendency of the spatial oscillation within the interval $2\sigma$;
we may expect the nontrivial structure around $\kappa=0.5$ if it exists.

\begin{figure}[t]
 \centering
 \includegraphics[width=0.235\textwidth]{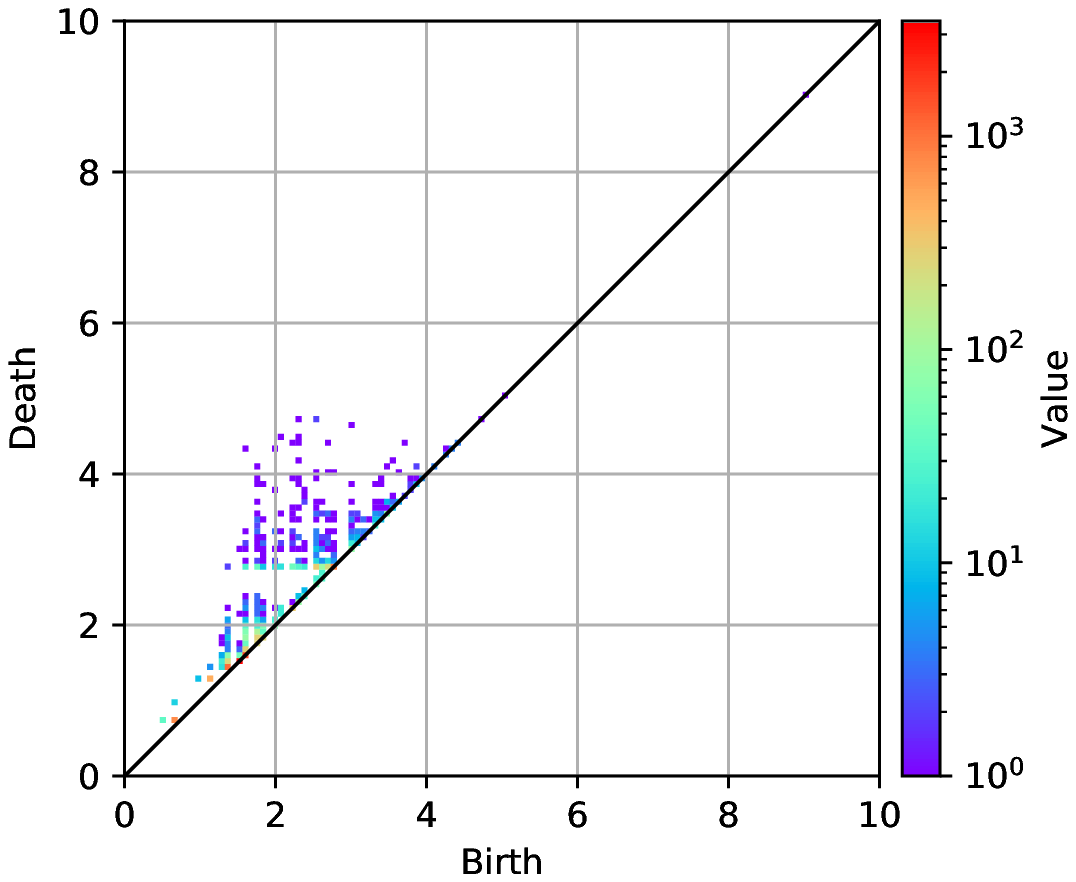}
 \includegraphics[width=0.235\textwidth]{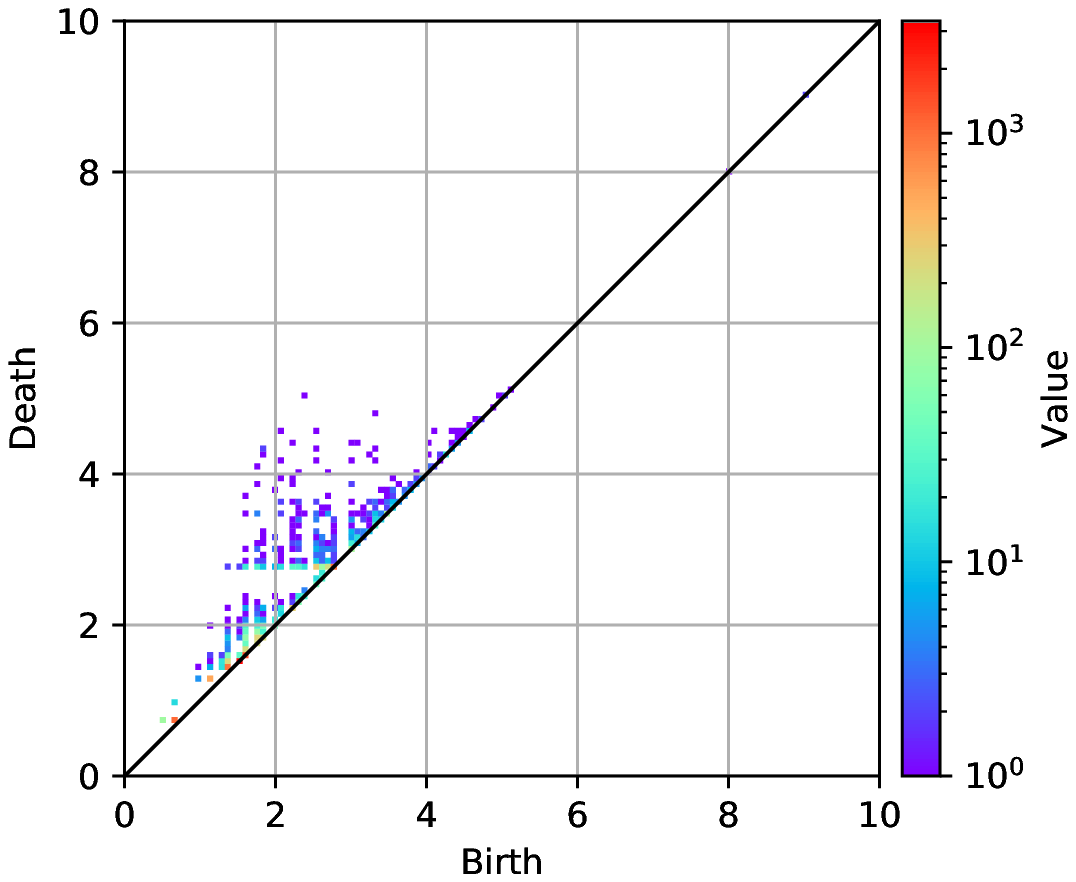}
 \includegraphics[width=0.235\textwidth]{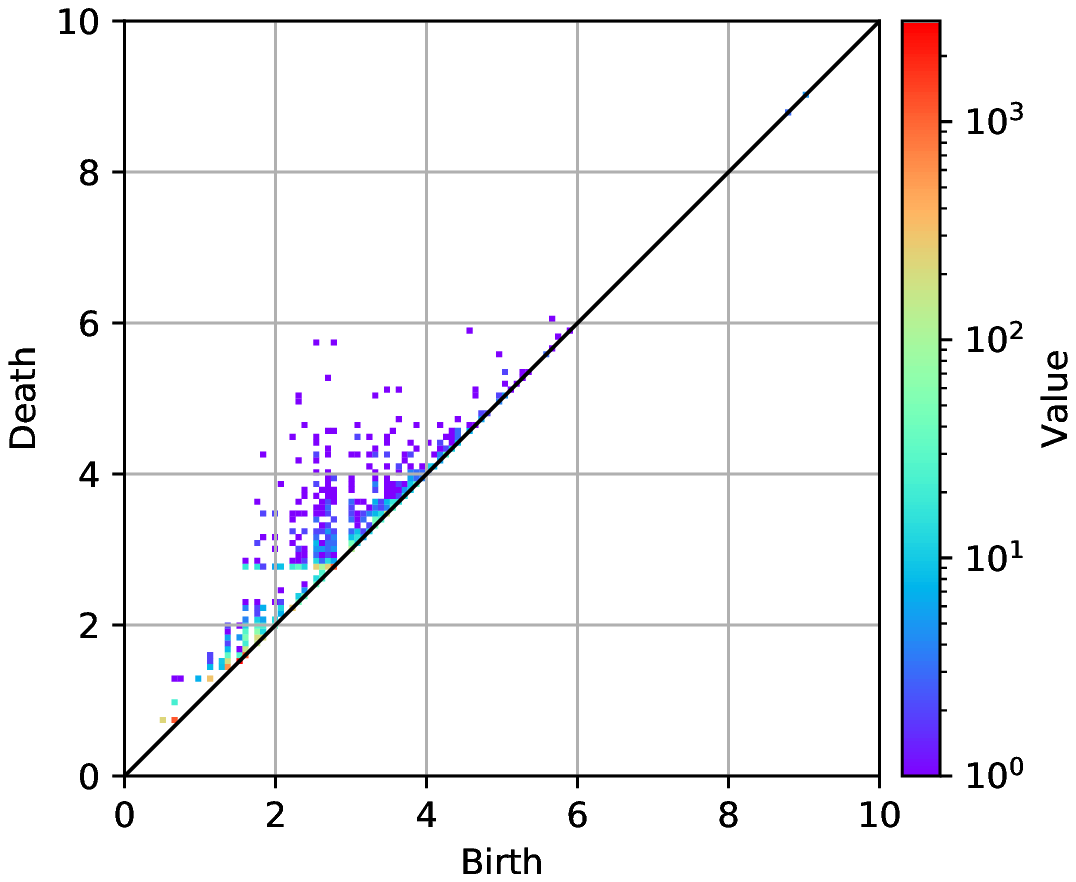}
 \includegraphics[width=0.235\textwidth]{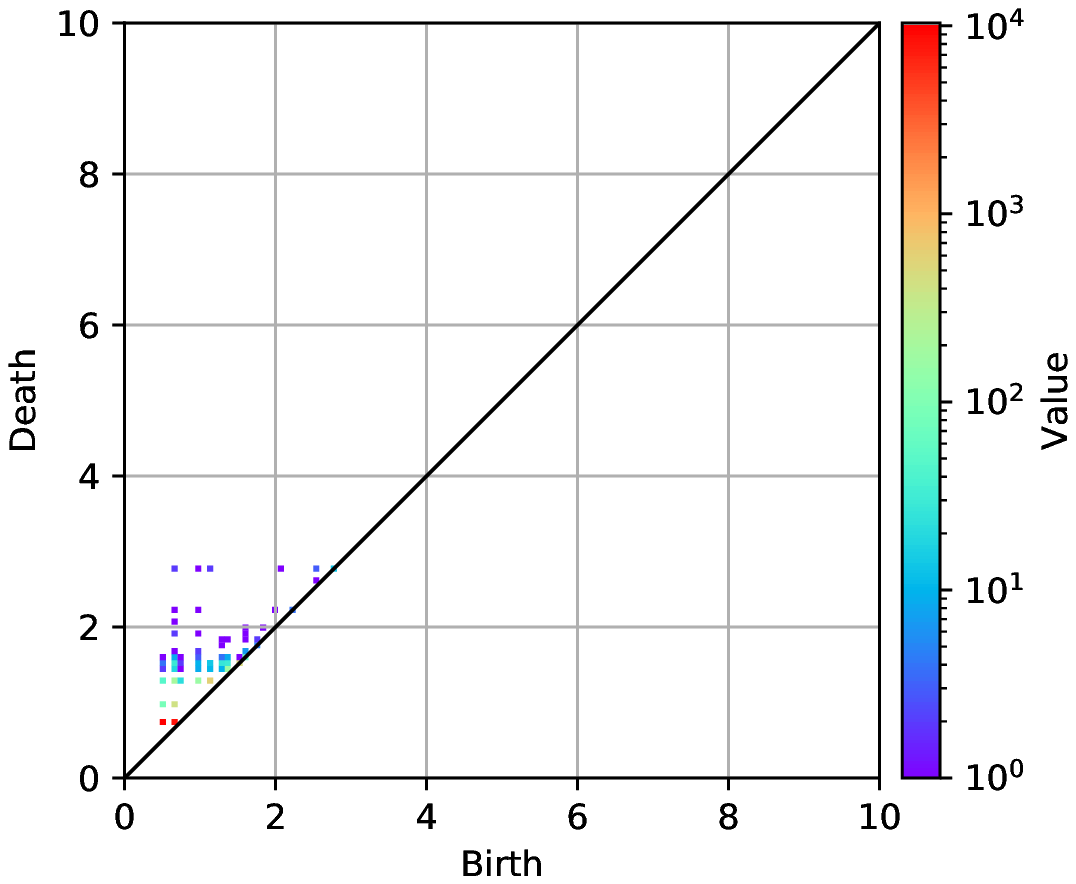}
 \caption{
 The persistent diagram at $\mu=0$ for $k=0$ spins; e.g., the data set A. Panels are results with $\kappa=0.3,0.4,0.5$ and $0.6$ for one particular configuration from the left-top $\to$ right-top $\to$ left-bottom $\to$ the right-bottom panels, respectively.
 }
\label{fig:per_k}
\end{figure}

\begin{figure}[t]
 \centering
 \includegraphics[width=0.34\textwidth]{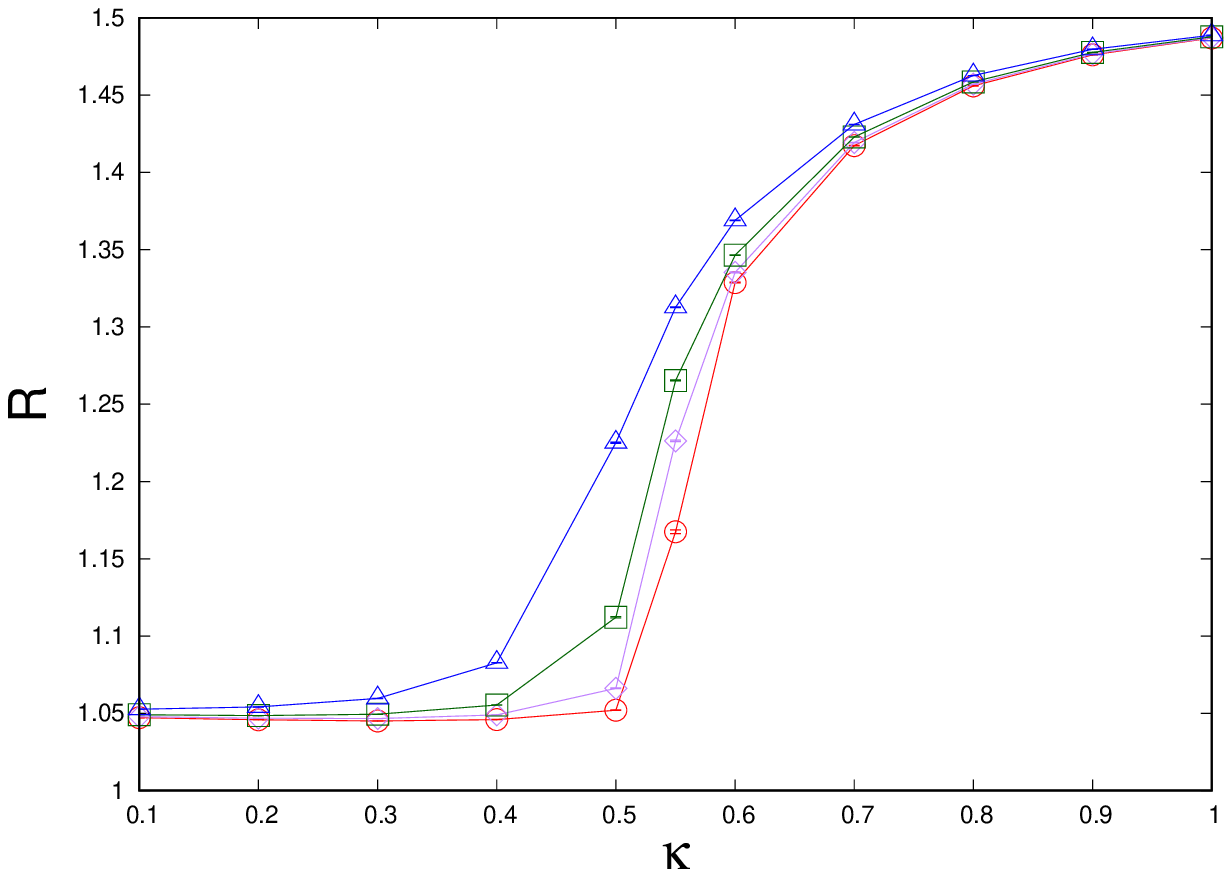}
 \includegraphics[width=0.34\textwidth]{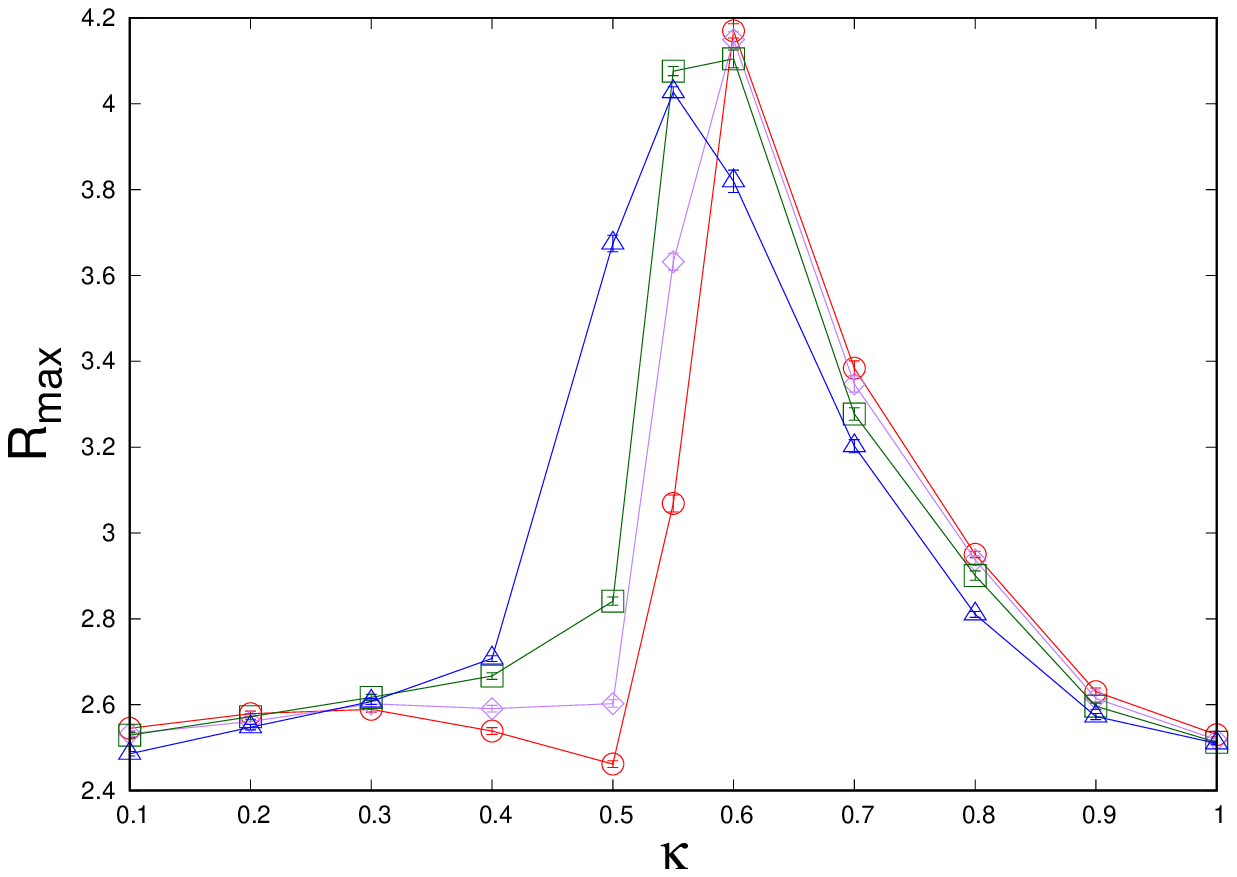}
 \caption{
 The $\kappa$-dependence of the mean value of the birth-death ratio with $\mu_\mathrm{iso}=0, 5, 6$ and $7$ where we take the configuration average.
 The top and bottom panels are the result of the averaged birth-death ratio ($R$) and the maximum birth-death ratio ($R_\mathrm{max}$), respectively.
 The open circle, diamond, square and triangle symbols are results with $\mu_\mathrm{iso}=0, 5, 6$ and $7$, respectively.
 Lines are just eye guides.
 }
\label{fig:ra_1}
\end{figure}

Figure~\ref{fig:per_k} shows the persistent diagrams at $\kappa=0.3, 0.4, 0.5$ and $0.6$ with $\mu_\mathrm{iso} = 0$.
The persistent homology analysis then detects the phase transition from sudden change of the persistent diagram; see also Fig.\,\ref{fig:ra_1}.
In addition, from the figures, we can see the qualitative difference at the intermediate $\kappa$.
When we compare results with $\kappa=0.3$ and $0.6$, the distribution seems to be simply shrank, but the distribution is temporally enlarged at intermediate $\kappa$.
This means that possible types of hole are changed and finally trivial cubes only persist.
This indicates that the system forms the cluster-like structure at intermediate $\kappa$ because QCD-like Potts model has the first-order thermal phase transition at $\mu_\mathrm{iso}=0$ in the thermodynamic limit.

\begin{figure}[t]
 \centering
 \includegraphics[width=0.235\textwidth]{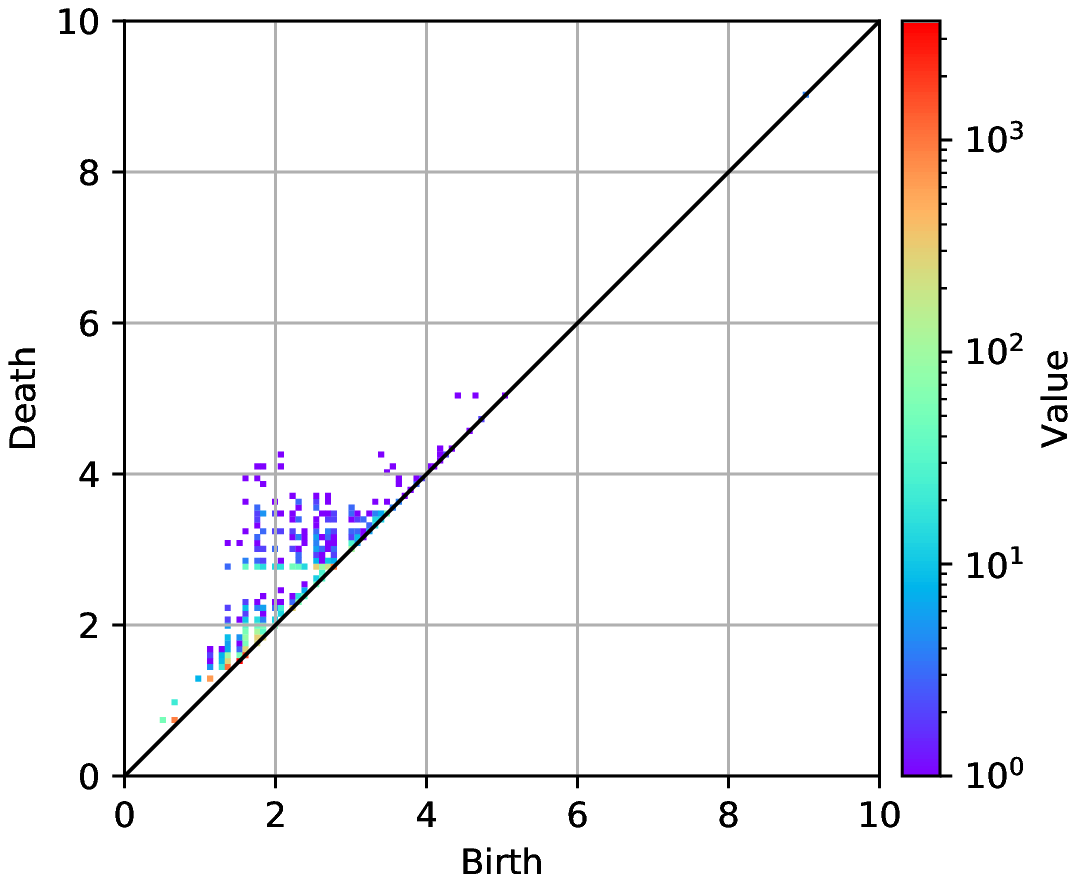}
 \includegraphics[width=0.235\textwidth]{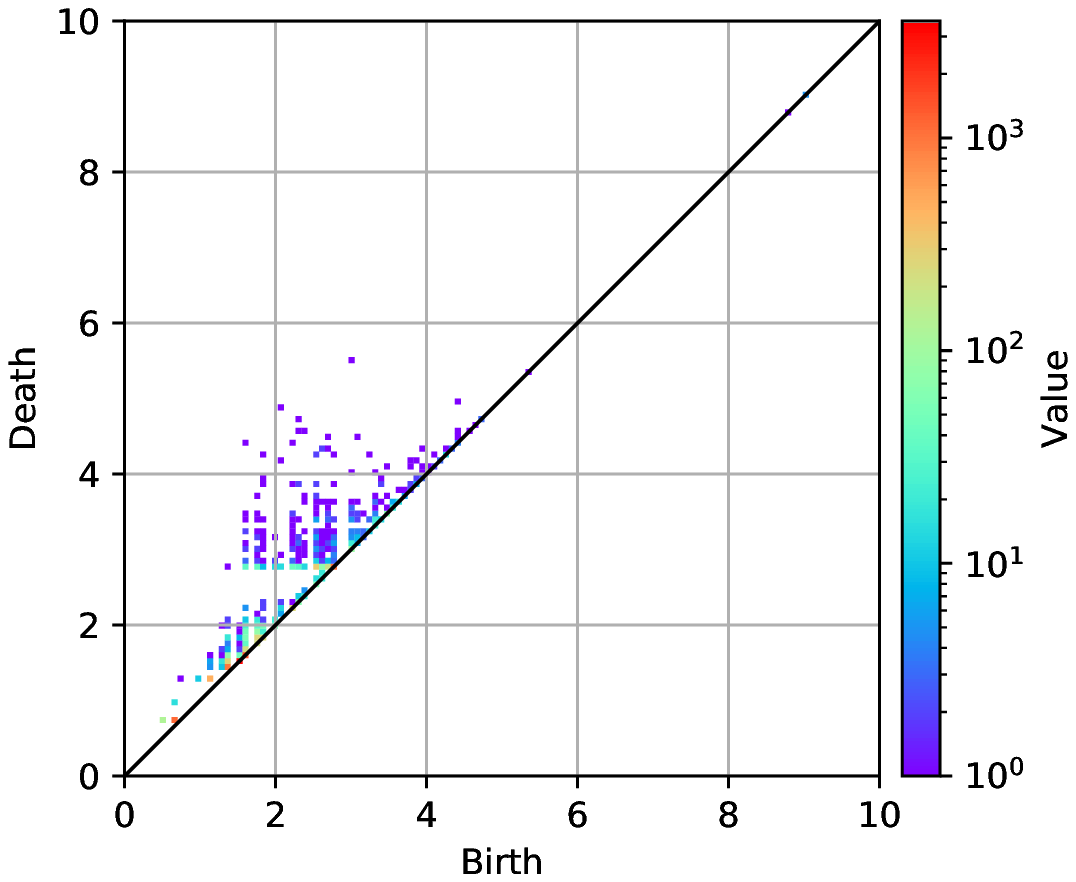}
 \includegraphics[width=0.235\textwidth]{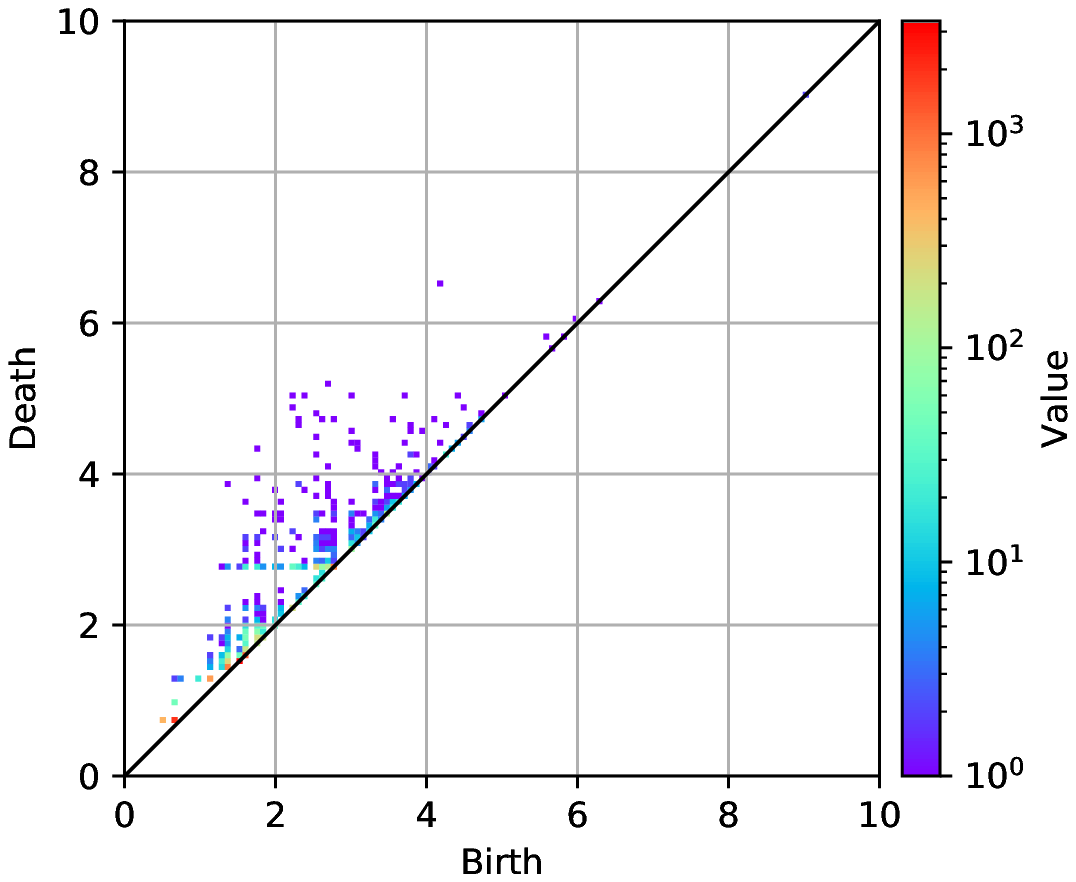}
 \includegraphics[width=0.235\textwidth]{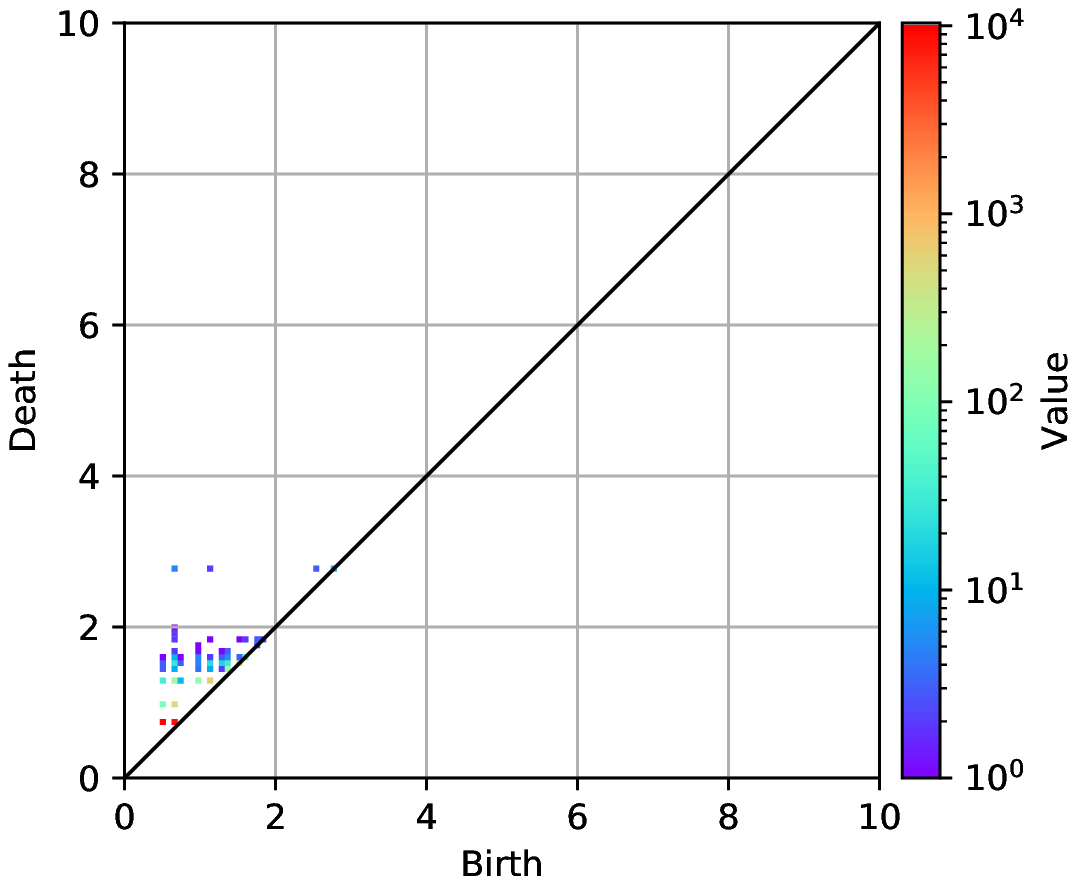}
 \caption{
 The persistent diagram at $\mu=5$ for $k=0$ spins; e.g., the data set A. Panels are results with $\kappa=0.3,0.4,0.5$ and $0.6$ for one particular configuration from the left-top $\to$ right-top $\to$ left-bottom $\to$ right-bottom panels, respectively.
 }
\label{fig:per_k_m6}
\end{figure}

Figure~\ref{fig:per_k_m6} shows the persistent diagrams at $\kappa=0.3, 0.4, 0.5$ and $0.6$ with $\mu_\mathrm{iso} = 5$ as an example of the intermediate $\mu_\mathrm{iso}$ case.
From these figures, we can see the similar tendency of Fig.\,\ref{fig:per_k} at intermediate $\kappa$, but we will see that the maximum birth-death ration is not changed so much unlike the result with $\mu_\mathrm{iso}=0$ as shown in Fig.\,\ref{fig:ra_1}; details are discussed in latter.

\begin{figure}[b]
 \centering
 \includegraphics[width=0.235\textwidth]{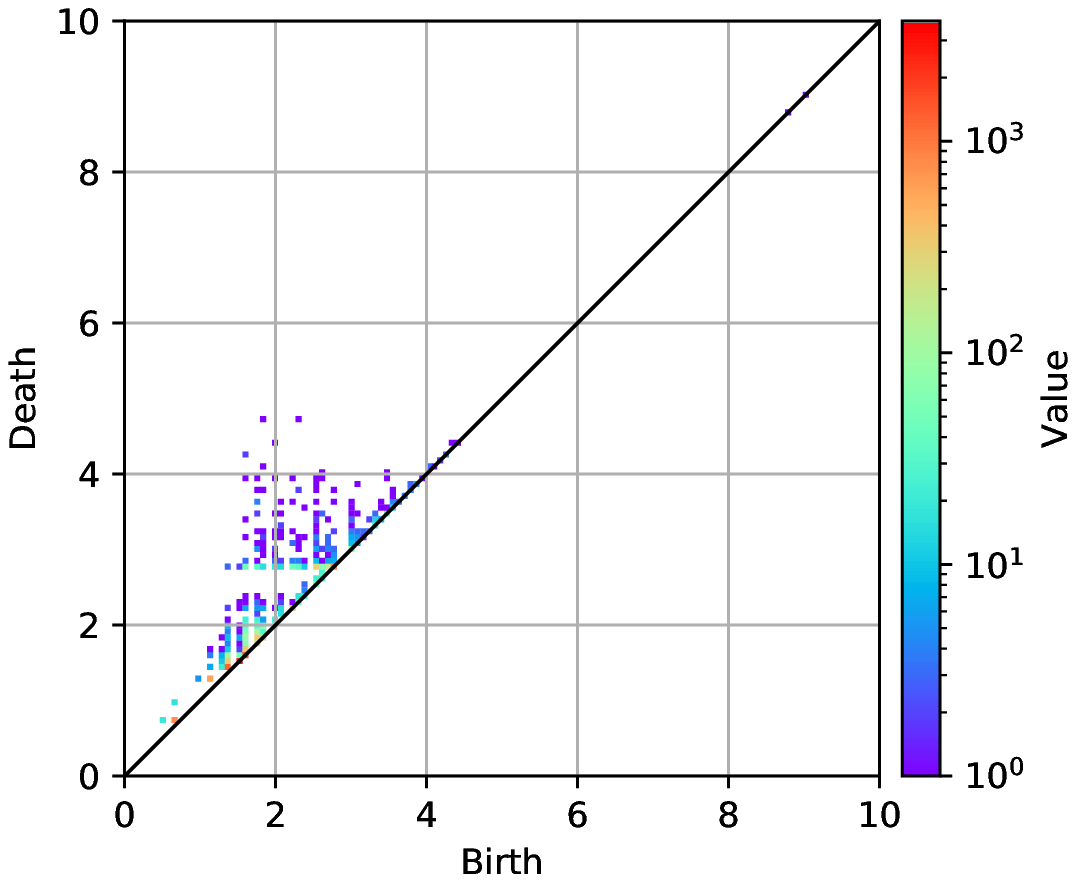}
 \includegraphics[width=0.235\textwidth]{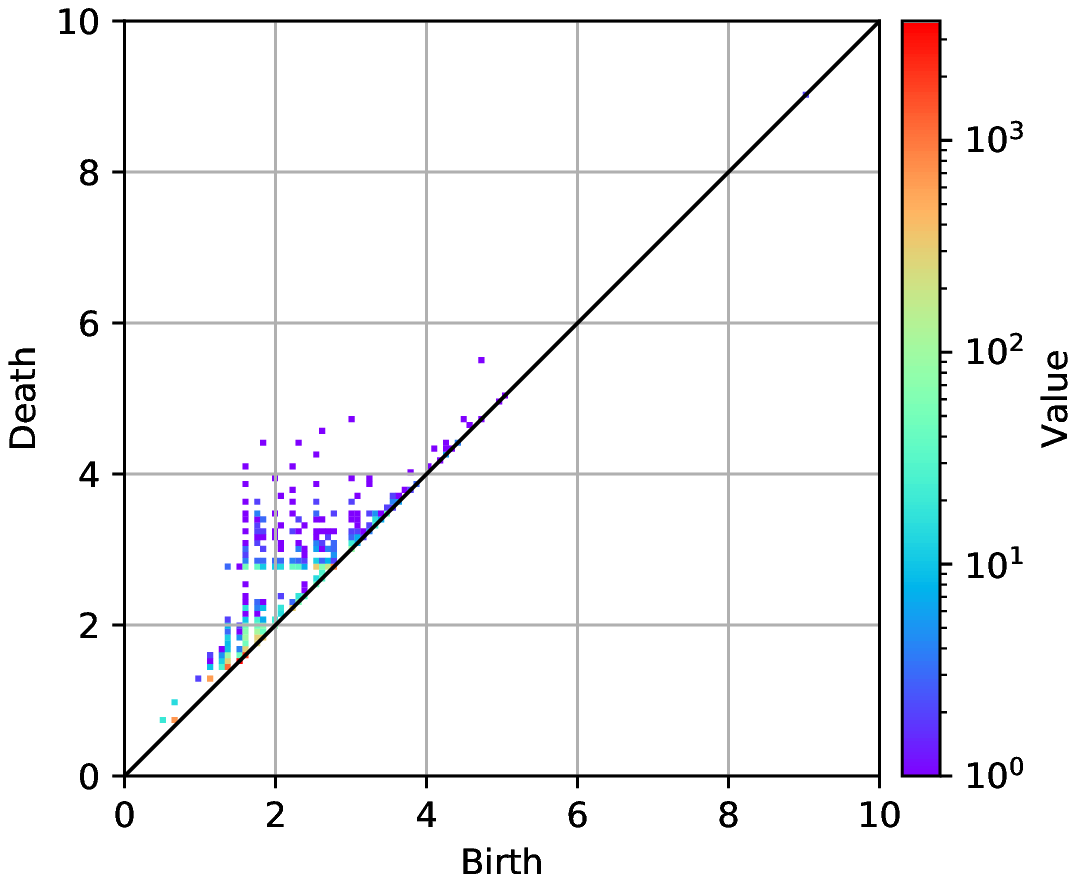}\\
 \includegraphics[width=0.235\textwidth]{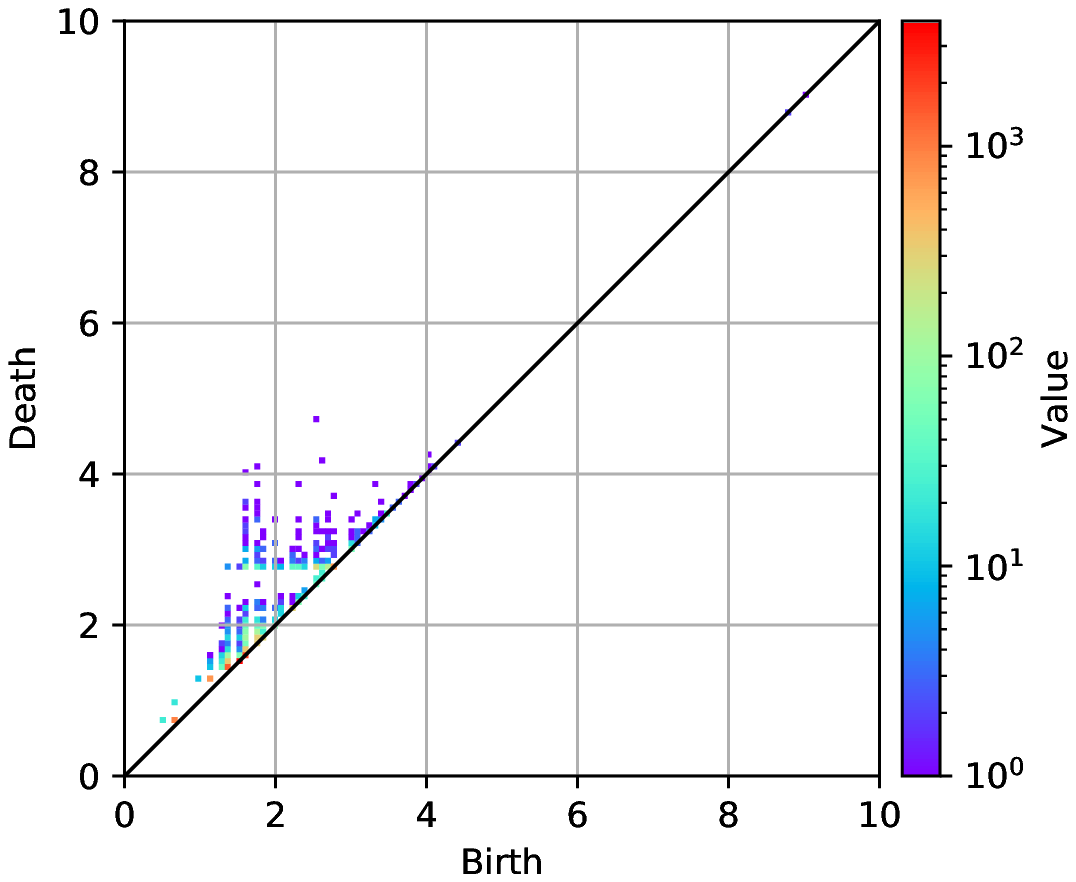}
 \includegraphics[width=0.235\textwidth]{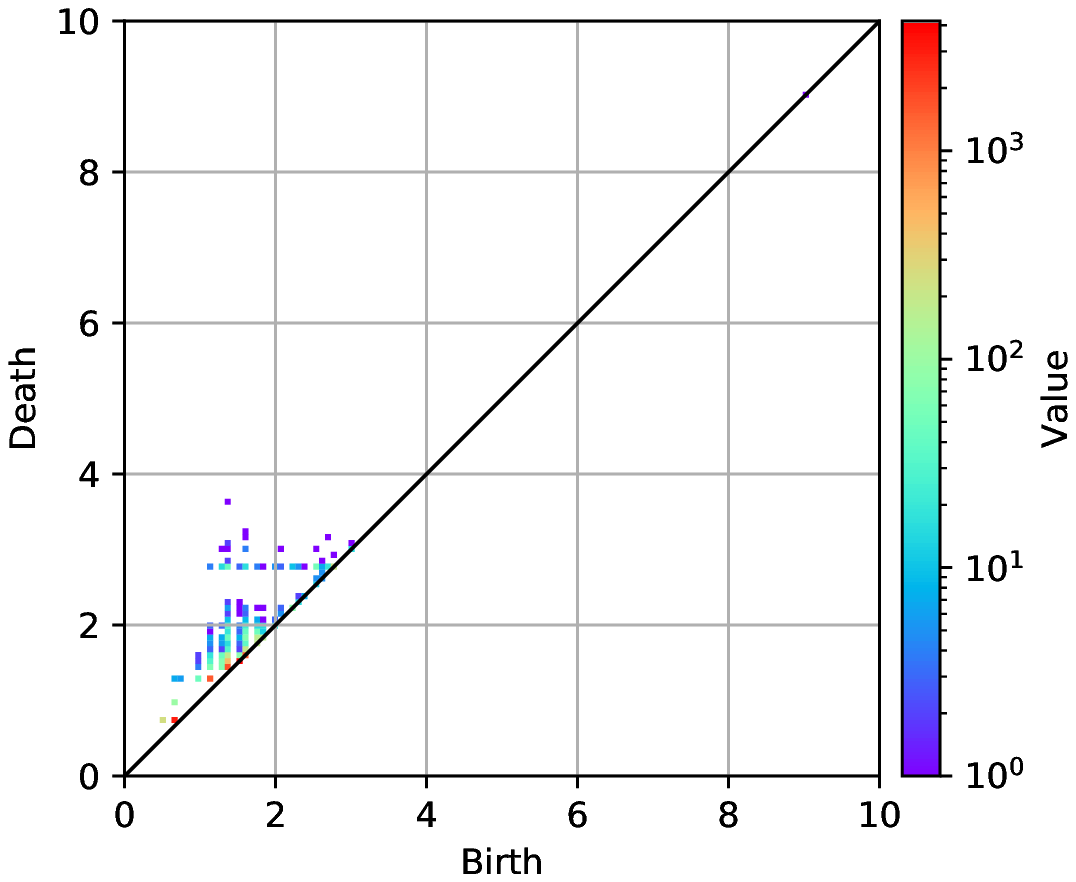}
 \caption{
 The persistent diagram at $\kappa=0.2$ for $k=0$ spins; e.g., the data set A. Panels are results with $\mu_\mathrm{iso}=2,4,6$ and $8$ for one particular configuration from the left-top $\to$ right-top $\to$ left-bottom $\to$ right-bottom panels, respectively.
 }
\label{fig:per}
\end{figure}

Figure~\ref{fig:per} shows the persistent diagrams at $\mu_\mathrm{iso}=2,4,6$ and $8$ with $\kappa=0.2$.
There are no nontrivial change at finite $\mu_\mathrm{iso}$.
This indicates that there are no first-order phase transition and also nontrivial spatial structure; we can expect that the crossover is realized along the $\mu_\mathrm{iso}$ direction at $\kappa=0.2$.
Actually, the chiral symmetry is not present in this model and thus there are no mechanisms which enhance the phase transition at low $\kappa$.
To investigate the persistent homology more deeply, we next consider the averaged ratio and the maximum ratio of the birth and the death time.

Above persistent diagrams are obtained by using one particular configuration and thus we here show the result of the ratio of the birth and death times with configuration average in Fig.\,\ref{fig:ra_1}.
Statistical errors are small and thus they are in the symbols.
The averaged ratio of the birth and death times are shown in the top panel of Fig.\,\ref{fig:ra_1}; we can expect that the averaged ratio is responsible to bulk properties of the system.
From the upper panel of the figure, we can see the clear tendency of the first-order transition at small $\mu_\mathrm{iso}$ from the steep change of the averaged ratio.
Since the holes which have distinct ratio from the diagonal line on the persistent diagram are responsible for important spatial structures in the persistent homology analysis, we also show the maximum ratio at $\mu_\mathrm{iso}=0,5,6$ and $7$.
At low $\mu_\mathrm{iso}$, the maximum ratio is temporally decreased at intermediate $\kappa$ and it has the peak after that its decreasing behavior is ended. 
This indicates that small spatial structures are developed when we approaches to the first-order transition and after larger spatial structures are formed.
Interestingly, we can find the flat region with increasing $\kappa$ at intermediate $\mu_\mathrm{iso} \sim 5$.
In this region, there are no change of the topological (large spatial) structure of the system and it may indicate the critical endpoint because the enhancement of the correlation length is expected and then the topological structure may not be changed so match; there may be the block-spin transformation invariance for the large spatial structure.
This also indicates that there are no nontrivial spatial structures such as the oscillation near the critical endpoint at least in this model.

\begin{figure}[t]
 \centering
 \includegraphics[width=0.38\textwidth]{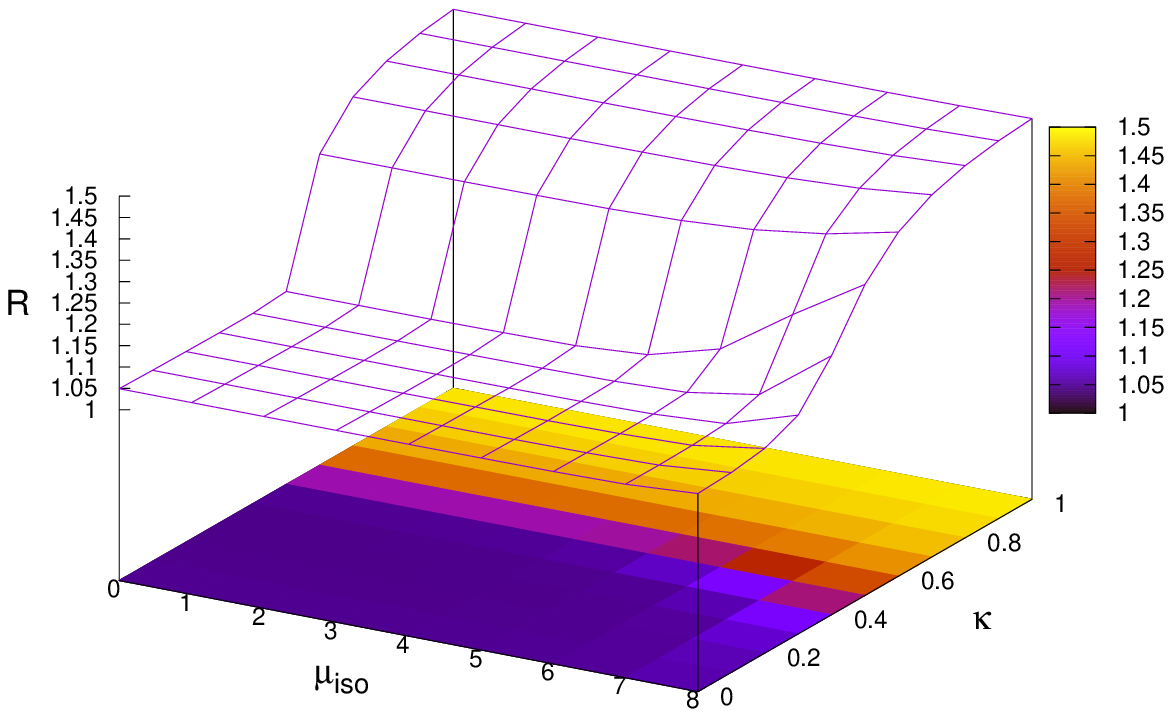}
 \includegraphics[width=0.38\textwidth]{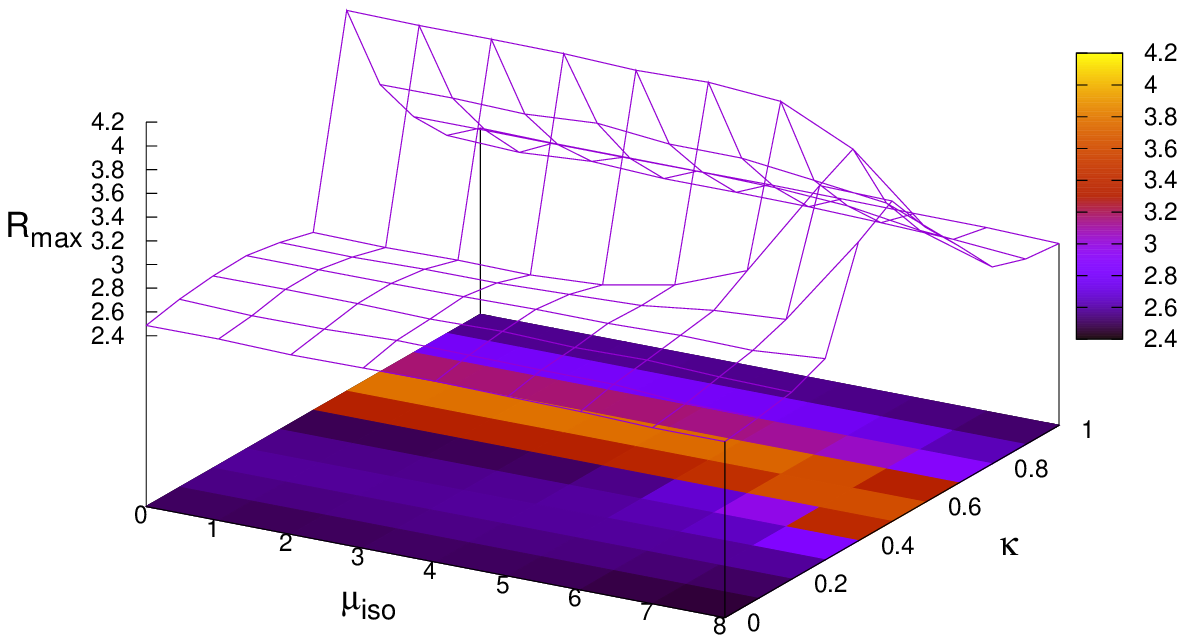}
 \caption{
 The mean value of the birth-death ratio on the $\mu_\mathrm{iso}$-$\kappa$ plane where we take the configuration average.
 The top and bottom panels are the result of the averaged birth-death ratio $R$ and the maximum $R_\mathrm{max}$, respectively.
 Statistical errors are very small and thus we do not show them here.
 }
\label{fig:ra_2}
\end{figure}

Finally, we show the ratio of the birth and death times for each $(\kappa,\mu_\mathrm{iso})$ in Fig.\,\ref{fig:ra_2}.
The top and the bottom panels show the averaged ratio ($R$) and the maximum ratio ($R_\mathrm{max}$), respectively.
From the panels, we can image the behaviors of the averaged ratio and the maximum ratio in the whole $(\mu_\mathrm{iso},\kappa)$ region.
The behavior of the averaged ratio is well matched with the Polyakov loop, but we can see that the maximum ratio has more information about the system.
The persistent homology, thus, has not only the information of the bulk properties of the system but also the spatial structure.

\section{Summary}
\label{Sec:summary}

In this paper, we have investigate the phase structure of the three-dimensional three-state Potts model at finite isospin chemical potential ($\mu_\mathrm{iso}$); we call it QCD-like Potts model.
Since the Potts model with the external field has the sign problem and thus we sidestep the sign problem by using the isospin chemical potential.
The spin configurations are then generated by using the QCD-like Potts model via the simple Metropolis algorithm, and the spatial structure of it is explored.
Even if the isospin chemical potential leads some differences to QCD-like Potts model comparing with original one, we can expect to mimic the correct spin configuration at finite density via the QCD-like Potts model.
Then, we have performed the persistent homology analysis to investigate the spatial structure of the spin configuration.

It has been obtained that the averaged ratio of the birth and death times is well matched with the behavior of the Polyakov loop and thus this quantity is only responsible to the bulk properties of the system.
On the other hand, the maximum ratio of the birth-death ratio is expected to be responsible to the spatial structure in addition to the bulk properties of the system.
We found that there may be no nontrivial spatial structure in the QCD-like Potts model at high and intermediate $\mu_\mathrm{iso}$; it can be seen from the behavior of the maximum ratio of the birth-death times.

In the case of realistic QCD, we do not have the spin configuration and thus we need some more extension to use the persistent homology.
However, we can have the gauge configuration in QCD and thus similar analysis can be possible; for example, we can classify each gauge field at each site from the viewpoint of the center symmetry structure of the Polyakov loop which has been employed in the investigation of the center clustering structure.
Of course, gauge configurations are more complicated quantity than the Potts spin and thus we need some more progresses for the analysis. 
We hope this study sheds new lights to exploring the phase structure of QCD at finite $\mu_\mathrm{R}$ from the topological viewpoint based on the persistent homology.

\begin{acknowledgments}
This work is supported in part by the Grants-in-Aid for Scientific Research from JSPS (No. 18K03618, 19H01898 and 20K03974).
\end{acknowledgments}

\appendix

\section{Problem of complxification}
\label{sec:ap1}

One can think why we do not use the complxification of dynamical variables in the present Potts model because the complexification approach is quite powerful approach to sidestep the sign problem in QCD.
In this section, first, we summarize the details of the complxification of dynamical variables in QCD and after we discuss the problem appearing in the Potts model.

\subsection{Complexification in QCD}

It is well known that the ${\cal CK}$ symmetry imposing on the Dirac operator can sidestep the sign problem at least in the mean-field level computation of the Nambu--Jona-Lasinio (NJL) type QCD effective model which includes the Polyakov-loop dynamics \,\cite{Nishimura:2014rxa,Nishimura:2014kla}.
Below, all discussions are done with the Polyakov-gauge fixing, $\partial_4 A_4 =0$, and $A_4$ is then diagonalized by using the remaining spatial gauge degree of freedoms.

Without imposing the ${\cal CK}$ symmetry, the temporal gluon field ($A^4$) are given by
\begin{align}
    A^4({\bf x}) &= A^4_3({\bf x}) \lambda_3 + A^4_8({\bf x}) \lambda_8,
    \label{Eq:QCD}
\end{align}
and the Polyakov loop ($\Phi$) then becomes
\begin{align}
    \Phi({\bf x}) &= \frac{1}{3} \Bigl( e^{i\beta \phi_1({\bf x})} + e^{i\beta \phi_2({\bf x})} + e^{i\beta \phi_3 ({\bf x})}  \Bigr),
    \label{Eq:QCD2}
\end{align}
where  $\lambda_3$ and $\lambda_8$ are diagonal components of the Gell-Man matrices, $\beta=1/T$ is the inverse temperature and
\begin{align}
    \phi_1({\bf x}) &= A_3({\bf x})+\frac{A_8({\bf x})}{\sqrt{3}},~~~~
    \phi_2({\bf x}) = -A_3({\bf x})+\frac{A_8({\bf x})}{\sqrt{3}},
    \nonumber\\
    \phi_3({\bf x}) &= -\frac{2A_8({\bf x})}{\sqrt{3}}.
\label{Eq:A4}
\end{align}
here $A_3, A_8 \in \mathbb{R}$.

When we introduce the complexification of dynamical variables, the temporal component of gluon field becomes
\begin{align}
    A^4({\bf x}) &= [A^{4,\mathrm{R}}_3 ({\bf x}) + i A^{4,\mathrm{I}}_3 ({\bf x})]  \lambda_3 
    \nonumber\\
    &+ [A_8^{4,\mathrm{R}}({\bf x}) + i A^{4,\mathrm{R}}_8({\bf x})] \lambda_8,
    \label{A4_comp}
\end{align}
where $A^{\mathrm{R,I}}_{3,8} \in \mathbb{R}$.
The ${\cal CK}$ symmetry imposing is corresponding to the setting with $A^{4,\mathrm{I}}_3=A^{4,\mathrm{R}}_8=0$. 
This modification means that we replace the real $A_8$ with pure imaginary one.
This ${\cal CK}$ symmetric choice, Eq.\,(\ref{A4_comp}) with $A^{8,\mathrm{I}}_3=0$ and $A^{8,\mathrm{R}}_8=0$, is the special case of the complexified dynamical variables approach.
Of course, we can also complexify the third component of the gluon field $A_3$ if we want.

It should be noted that the ${\cal CK}$ symmetry realization in the QCD effective model which includes the Polyakov-loop dynamics such as the Polyakov-loop extended NJL (PNJL) model~\cite{Fukushima:2003fw} is mathematically and numerically proven by using the Lefschetz thimble method~\cite{Tanizaki:2015pua} and the path optimization method~\cite{Kashiwa:2018vxr} at least for the homogeneous solution.
Although we do not know how the symmetry efficiently weaken the sign problem in realistic QCD including quantum fluctuation, we can expect that the symmetry weaken the sign problem.

\subsection{Complexification in Potts model}
\label{Sec:Potts2}
To introduce complexification of dynamical variables to the Potts model, we encounter the serious problem since there are too much simplifications on the gauge field treatment in the Potts model.

The Polyakov loop in the Potts model with the complexification of dynamical variables is defined as
\begin{align}
 \Phi({\bf x})
 &= \frac{1}{3} \Bigl(
      e^{i\beta \tilde{\phi}_{1,{\bf x}}}
    + e^{i\beta \tilde{\phi}_{2,{\bf x}}}
    + e^{i\beta \tilde{\phi}_{3,{\bf x}}}  \Bigr),    
    \nonumber\\
 {\bar \Phi}({\bf x})
 &= \frac{1}{3} \Bigl(
      e^{-i\beta \tilde{\phi}_{1.{\bf x}}}
    + e^{-i\beta \tilde{\phi}_{2,{\bf x}}}
    + e^{-i\beta \tilde{\phi}_{3,{\bf x}}}  \Bigr),    
\end{align}
where
\begin{align}
    \tilde{\phi}_{1,{\bf x}}
    &= A^{4,\mathrm{R}}_{3,{\bf x}} + i A^{4,\mathrm{I}}_{3,{\bf x}} +\frac{A^{4,\mathrm{R}}_8({\bf x})+i A^{4,\mathrm{I}}_{8,{\bf x}}}{\sqrt{3}},
    \nonumber\\
    \tilde{\phi}_{2,{\bf x}}
    &= -[A^{4,\mathrm{R}}_{3,{\bf x}} + i A^{4,\mathrm{I}}_{3,{\bf x}}] +\frac{A^{4,\mathrm{R}}_{8,{\bf x}}+i A^{4,\mathrm{I}}_{8,{\bf x}}}{\sqrt{3}},
    \nonumber\\
    \tilde{\phi}_{3,{\bf x}} &= -\frac{2[A^{4,\mathrm{R}}_{8,{\bf x}} + i A^{4,\mathrm{I}}_{8,{\bf x}}]}{\sqrt{3}},
\end{align}
here the Potts model requires
\begin{align}
    \beta \frac{A^{4,\mathrm{R}}_{8,{\bf x}}}{\sqrt{3}} = \frac{2 \pi k}{3}.
    \label{Eq:A_8}
\end{align}
To make the Potts energy real by using $A_3^{4,\mathrm{R}}$, $A_3^{4,\mathrm{I}}$ and $A_8^{4,\mathrm{I}}$, we can find one possible setting from the structure of the Dirac operator in QCD;
the actual possible form is
\begin{align}
    A_3^{4,\mathrm{R}} = A_8^{4,\mathrm{R}},~~
    A_3^{4,\mathrm{I}} = -\frac{3}{2}\mu,~~
    A_8^{4,\mathrm{I}} = \frac{\sqrt{3}}{2}\mu.
\end{align}
With this setting, we can easily check that the Dirac operator is real.
Then, $\Phi$ and ${\bar \Phi}$ becomes
\begin{align}
\Phi({\bf x})
 &= \frac{e^{\mu/T}}{3} \Bigl(
      e^{ 2\pi i k/3}
    + e^{-3 \mu/T}
    + e^{ 4\pi i k/3} \Bigr),
    \nonumber\\
{\bar \Phi}({\bf x})
 &= \frac{e^{-\mu/T}}{3} \Bigl(
      e^{ 4\pi i k/3}
    + e^{ 3 \mu/T}
    + e^{ 2\pi i k/3} \Bigr).
\label{Eq:extended}
\end{align}
In this setting, all additional dynamical variables are fixed by using $\beta \mu$.
Unfortunately, above Polyakov loop and its conjugate take $\frac{1}{3} \le \langle \Phi \rangle$ and it is not matched with those behaviors in QCD even at $\mu=0$.
Since $A_{8,{\bf x}}^{4,\mathrm{R}}$ takes the specific values (\ref{Eq:A_8}) in the Potts model and thus we can not suitably remove the sign problem via the complexification unlike the PNJL model: In the PNJL model case, we complexify $A_8^4$, but $A_3^4$ is kept as real value.

\bibliography{ref.bib}

\end{document}